\documentclass[prb, twocolumn, showkeys,showpacs,amsmath,amsfonts,floatfix,superscriptaddress,nofootinbib,longbibliography, 10pt, aps]{revtex4-2}
\usepackage[normalem]{ulem}

\usepackage{amssymb,amsmath,amstext}                
\usepackage{graphicx}   
\usepackage{lineno}
\usepackage{epstopdf}                                             
\usepackage{color}       
\usepackage{bm}
\usepackage{appendix}       
\usepackage{latexsym}
\usepackage[version=4]{mhchem}
\usepackage[colorlinks=true,citecolor=blue,linkcolor=magenta]{hyperref}
\usepackage{nameref} 
\usepackage{xcolor}

\def\be{\begin{equation}}
\def\ee{\end{equation}}
\def\bea{\begin{eqnarray}}
\def\eea{\end{eqnarray}}
\def\bi{\begin{itemize}}
\def\ei{\end{itemize}}

\newcommand{\ket}[1]{\mbox{$| #1 \rangle$}}

\makeatletter
\def\moverlay{\mathpalette\mov@rlay}
\def\mov@rlay#1#2{\leavevmode\vtop{%
   \baselineskip\z@skip \lineskiplimit-\maxdimen
   \ialign{\hfil$\m@th#1##$\hfil\cr#2\crcr}}}
\newcommand{\charfusion}[3][\mathord]{
    #1{\ifx#1\mathop\vphantom{#2}\fi
        \mathpalette\mov@rlay{#2\cr#3}
      }
    \ifx#1\mathop\expandafter\displaylimits\fi}
\makeatother

\begin{document}

\title{Forestalled Phase Separation as the Precursor to Stripe Order}

\author{Aritra Sinha} 
\email{aritrasinha98@gmail.com}
\affiliation{Max Planck Institute for the Physics of Complex Systems, 
             N\"{o}thnitzer Strasse 38, Dresden 01187, Germany}
             
\author{Alexander Wietek} 
\affiliation{Max Planck Institute for the Physics of Complex Systems, 
             N\"{o}thnitzer Strasse 38, Dresden 01187, Germany}

\date{\today}

\begin{abstract}

Stripe order is a prominent feature in the phase diagram of the high-temperature cuprate superconductors. It has been confirmed as the lowest-energy state of the two-dimensional Fermi Hubbard model in certain parameter regimes. Upon increasing the temperature, stripes and the superconducting state give way to the enigmatic strange-metal and pseudogap regimes, whose precise nature remains a long-standing puzzle. Using modern tensor network techniques, we discover a crucial aspect of these regimes. Infinite projected entangled pair states (iPEPS) simulations in the fully two-dimensional limit reveal a maximum in the charge susceptibility at temperatures above the stripe order. This maximum is located around filling $n=0.9$ and intensifies upon cooling. Using minimally entangled typical thermal states (METTS) on finite cylinders, we attribute the enhanced charge susceptibility to the formation of charge clusters, reminiscent of phase separation where the system is partitioned into hole-rich and hole-depleted regions. In contrast to genuine phase separation, the charge cluster sizes fluctuate without a divergent charge susceptibility. Hence, while this precursor state features clustering of charge carriers, true phase separation is ultimately forestalled at lower temperatures by the onset of stripe order.

\end{abstract}

\maketitle


\noindent Phase separation (PS) in high-$T_{\text c}$ superconductors (HTSCs) refers to an underlying tendency toward coexistence of hole-rich metallic regions and hole-poor antiferromagnetic domains~\cite{bookPS, dagotto1994}. Experimentally, such inhomogeneities were observed early on in \ce{La2CuO_{4+\delta}}~\cite{grenier1992phase} and later in stripe-ordered \ce{La_{1.6-x}Nd_{0.4}Sr_{x}CuO_{4}}~\cite{tranquada1995evidence}. Beyond cuprates~\cite{pan2001microscopic, mcelroy2005atomic, campi2015inhomogeneity}, PS tendencies also appear in iron-based chalcogenides~\cite{yuan2012nanoscale, ricci2011}, doped manganites~\cite{miao2020direct}, and nickelates~\cite{campi2022nanoscale}; see Ref.~\cite{kagan2021electronic} for a recent review.

Theoretical scenarios link PS, charge-density modulations, and superconductivity~\cite{bookPS, dagotto1994}. Early mean-field theories proposed that doped carriers cluster into metallic networks pivotal for superconductivity~\cite{bookPS, hizhnyakov1988high, hizhnyakov1989percolation}. Concepts of charge density waves (CDW) and stripe order~\cite{zaanen1989, Poilblanc1989, machida1989magnetism, zaanen1996striped} emerged in tandem with ideas about electronic inhomogeneities~\cite{caprara2019ancient}. Emery and Kivelson’s frustrated PS scenario~\cite{emery1993frustrated} posited that long-range Coulomb interactions could prevent full macroscopic segregation, favoring CDW-like patterns. Recent work on Hubbard-Holstein and related models~\cite{takahiro2017,julia2020nanoscale, karakuzu2022stripe} further underscores the importance of electron-lattice interactions in facilitating phase separation.

The square-lattice two-dimensional (2D) Fermi-Hubbard model (FHM)\cite{hubbard1963electron} is central to understanding high-Tc superconductivity\cite{anderson1987resonating,zhang1988,emery1980s}. Its phase diagram, restricted to nearest-neighbor (NN) hopping, has been debated extensively~\cite{argumenthellberg, argumentwhite}, but a growing consensus points to stripe order at slight doping under cuprate-relevant conditions~\cite{zheng2017stripe, steven2003, hager2005, zheng2016, huang2017numerical, huang2018stripe, ehlers2017, qin2020, jiang2020}. The pseudogap and strange-metal regimes, prominent in cuprates above the stripe order and superconducting phases, have driven intense scrutiny of the Hubbard model at finite temperature~\cite{huscroft2001,kyung2006, gull2013,huang2019strange, meixner2024, simkovic2024}. Notably, a cellular dynamical mean-field theory (CDMFT) study~\cite{kyung2006} tied pseudogap formation to short-range spin correlations, and diagrammatic Monte Carlo simulations~\cite{simkovic2024} suggest a predominantly spin-driven origin. Another study~\cite{meixner2024} using CDMFT reinforced the spin-driven origin of the pseudogap, while highlighting that charge inhomogeneities may play some role. 

Here, we study the FHM at finite temperature and slight doping regimes where pseudogap and strange-metal behaviors arise. Using two advanced tensor network methods—minimally entangled typical thermal states (METTS)\cite{white2009, stoudenmire2010minimally} using matrix product states (MPS)\cite{fannes1992finitely, ostlund1997}, and purification~\cite{verstraete2004matrix} with infinite projected entangled pair states (iPEPS)~\cite{jordan2008classical, corboz2010}—we examine both quasi-1D cylinders and infinite 2D systems. Our simulations reveal an incipient charge clustering promoted by antiferromagnetic (AFM) correlations at intermediate temperatures, suggesting an incipient PS, which is ultimately preempted by stripe order at lower temperatures.

\vspace{0.3cm}
\noindent{\large \textbf{Results}}
\vspace{0.2cm}

\noindent \textbf{Charge susceptibility in the thermodynamic limit} 

\noindent In this section, we implement the purification method~\cite{czarnik2019} via fermionic iPEPS~\cite{corboz2010} following Ref.~\cite{sinha2022}. Here, we use the particle-hole symmetric form of the FHM,
\bea
H &=& 
- \sum_{\langle i,j\rangle, \sigma} 
  t\left( c_{i\sigma}^\dag c_{j\sigma} + c_{j\sigma}^\dag c_{i\sigma} \right) + \nonumber\\
  & &
  \sum_i U \left( n_{i\uparrow} - \frac12 \right) \left( n_{i\downarrow} - \frac12 \right) -
  \sum_i \mu \; n_i.
\label{H}
\eea

\begin{figure*}[t!]
\vspace{-1cm}
\includegraphics[width=0.99\textwidth,clip=true]{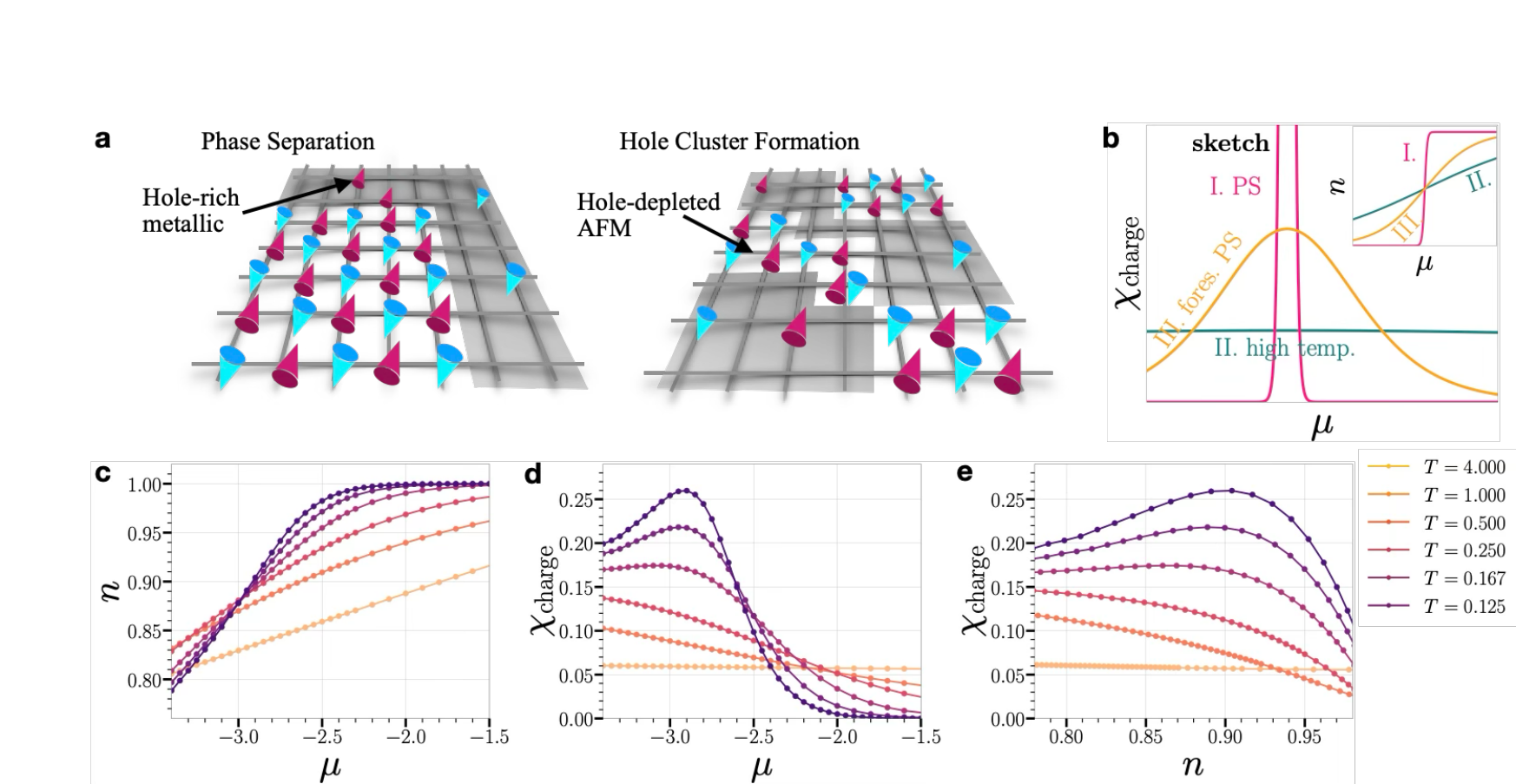}
\vspace{-0.3cm}
\caption{\textbf{Charge susceptibility. a.} Cartoon illustrations of phase separation in the Fermi-Hubbard model on a square lattice. The left panel shows complete phase separation, with a hole-rich phase (shaded grey) and an antiferromagnetic domain (unshaded), with red and blue cones representing spin-up and spin-down polarization of the electrons. The right panel depicts hole clustering, where thermal fluctuations prevent complete phase separation, resulting in clusters of holes accompanied by AFM domains. \textbf{b.} Schematic of charge susceptibility $\chi_\text{charge} = \frac{\partial n}{\partial \mu}$ vs. chemical potential $\mu$, depicting three possible scenarios: \textbf{I} (pink curve) indicates sharp phase separation at the ground state, with a discontinuity in $n$ (inset) and a divergent peak in $\chi_\text{charge}$; \textbf{II} (green curve) represents high-temperature behavior with smoothed $n$ vs. $\mu$ and flat $\chi_\text{charge}$; and \textbf{III} (orange curve) displays near-critical behavior at finite temperature, with a small peak in $\chi_\text{charge}$. We refer to this scenario as forestalled phase separation (labeled `fores. PS' in the sketch). The bottom panel shows iPEPS simulations done on an infinite square lattice at $U=10$. In \textbf{c.}, we plot $n(\mu)$ vs $\mu$, and in \textbf{d.}, we show $\chi_{\text{charge}}$ as a function of $\mu$. We observe a shift from Scenario \textbf{II} to \textbf{III} for the latter as the temperature $T$ decreases from $T=4.000$ to $T=0.125$. Additionally in \textbf{e.}, we plot $\chi_{\text{charge}}$ vs. $n$ which highlights the values of the density where $\chi_{\text{charge}}$ is maximal.}
\label{fig:1}
\end{figure*}
\begin{figure*}[t!]
\vspace{-0cm}
\includegraphics[width=1.0\textwidth,clip=true]{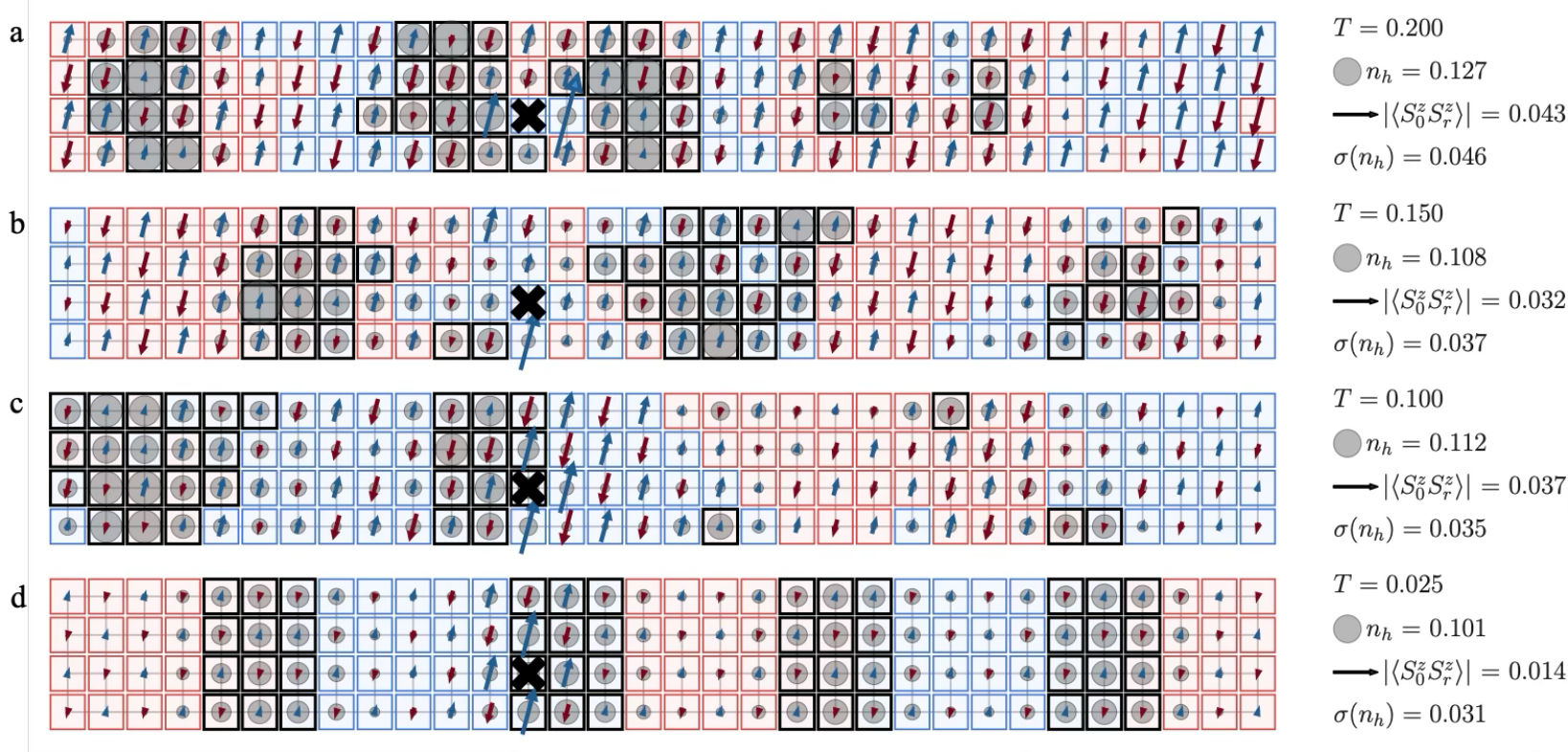}
\vspace{-0.3cm}
\caption{\textbf{METTS snapshots.} Visualizations of hole density and spin correlations for a typical METTS state $|\psi_i\rangle$ in the Hubbard model with $U=10$ at electron density $n=0.9375$ on a $32\times4$ cylinder. Grey circles, with diameters proportional to hole densities $n_{h}(r) = 1-\langle n_{r}\rangle=1-\langle\psi_i|n_r|\psi_i\rangle$, highlight variations in local charge distribution. Spin correlations are shown by arrows (red for negative and blue for positive) of lengths proportional to the correlation strengths $\langle\psi_i|S^{z}_0\cdot S^{z}_r|\psi_i\rangle$; the black cross denotes the origin $r=0$ of the spin correlator. Red and blue squares mark the sign and spatial structure of staggered spin correlations, $(-1)^{x+y}\langle\psi_i|S^{z}_0\cdot S^{z}_r|\psi_i\rangle$, thereby assigning the same color to each site within a given antiferromagnetic domain. Black-bordered sites indicate hole densities exceeding a specified threshold (see Eq.~(\ref{threshold})); neighboring black-bordered sites define the clusters $\mathcal{C}$ in Eq.~(\ref{clusters}). Panels (a)–(d), from top to bottom, show snapshots at temperatures $T=0.200$, $0.150$, $0.100$, and $0.025$, respectively. The right-hand scale uses fixed-size icons (constant reference circle radius and arrow length across all panels) for visual comparison and $\sigma_{n_h}$, the standard deviation of $n_{h}$ within a snapshot. At the lowest temperature ($T=0.025$), panel (d) shows stripe pattern, i.e., a charge-density wave intertwined with a spin-density wave. The antiferromagnetic correlations undergo a $\pi$-phase shift across the domain wall that runs through the middle of clusters of dominant size $m=|\mathcal{C}|=12$ sites.}

\label{fig:2}
\end{figure*}
\noindent where $t$ represents the NN hopping amplitude and $U > 0$ denotes the on-site Coulomb repulsion. Here, $c^{\dagger}_{i\sigma}$ ($c_{i\sigma}$) is the creation (annihilation) operator for an electron with spin $\sigma$ at site $i$, and $n_{i\sigma} = c^{\dagger}_{i\sigma} c_{i\sigma}$ is the number operator for electrons of spin $\sigma$ at site $i$. The summation $\langle i,j \rangle$ runs over NN sites on the square lattice. With this convention half-filling sits at $\mu=0$. In this article, we set $t=1$. We control the filling $n$ by changing the chemical potential $\mu$. In this grand-canonical formalism, the system exchanges particles with a reservoir, and the particle density $n$ adjusts to reach equilibrium, leading to a unique equilibrium density for a given chemical potential. The charge susceptibility $\chi_\text{charge} = \frac{\partial n}{\partial \mu}$ quantifies how sensitive the particle density is to changes in the chemical potential. The sketch Fig.~\ref{fig:1}(b) illustrates three scenarios with $\chi_\text{charge}$ versus $\mu$ plots  and density $n$ versus $\mu$ plots (inset). Scenario \textbf{I} (pink curve) shows PS at ground state for some $\mu = \mu_c$ with a sharp discontinuity in the $n$ vs $\mu$ plot and a divergent peak in $\chi_\text{charge}$, suggesting enhanced 
susceptibility. This signature has been used to provide evidence for PS in iPEPS studies of fermions on square and honeycomb lattices~\cite{corboz2010, Corboz_2012}. Because the iPEPS unit cell selected here is small and translationally invariant, actual inhomogeneous states cannot be represented; one can detect PS tendencies only via peaks in $\chi_\text{charge}$. Scenario \textbf{II} (green curve) describes a high-temperature state with thermal fluctuations that smooth the $n$ vs $\mu$ curve and keeps $\chi_\text{charge}$ 
relatively flat. There is also a scenario \textbf{III} (orange curve), which demonstrates a finite-temperature crossover. Here, the $n$ vs $\mu$ curve exhibits a noticeable change in curvature without a 
discontinuous jump. The charge susceptibility $\chi_\text{charge}$ in this case shows a modest peak, signifying an increased, but finite, susceptibility to variations in $\mu$. As we show below, in the Hubbard model, this can result from statistically fluctuating hole clusters that mimic phase separation on a smaller scale without achieving full separation. Cartoon illustrations of Scenario \textbf{I} and 
\textbf{III} for the Fermi-Hubbard model are presented in Fig.~\ref{fig:1}(a), with Scenario \textbf{I} depicted on the left and Scenario \textbf{III} on the right. In the bottom panel of Fig.~\ref{fig:1}, we show iPEPS results for an infinite square lattice at $U=10$ --- (c) plots the filling $n$ versus chemical potential $\mu$, while (d) displays the corresponding charge susceptibility. At high temperature ($T=4$), the system follows the smooth behavior of Scenario II (no PS). As $T$ is lowered, the $n(\mu)$ curves develop an increasingly sharp inflection, and $\chi_{\rm charge}(\mu)$ peaks become more pronounced, signaling a crossover towards Scenario III. In Fig.~\ref{fig:1}(e) we 
re-plot the same charge susceptibility data against the density $n$ instead of $\mu$ and find that the maximum occurs in a broad region surrounding density $n \approx 0.91$.
Importantly, plaquette CDMFT studies of the square-lattice Hubbard model at strong coupling~\cite{reymbaut2019pseudogap} show that the van Hove singularity (defined via the maximum in the local density of states at the Fermi level) appears at much larger doping than our observed susceptibility peak near doping $1-n \approx 0.09$ for comparable Coulomb strength $U$. 
This strongly suggests that our result does not arise from a non-interacting band-structure effect, but instead reflects genuine strong-coupling physics associated with Mottness. This behavior is consistent with findings from dynamical cluster approximation (DCA)~\cite{macridin2006,khatami2010} for the Hubbard model with nearest-neighbor hopping only ($t' = 0$), which show that $\chi_\text{charge}$ stays finite at nonzero temperature yet increases as $T \rightarrow 0$. With a finite $t'$, however, the model was shown to undergo a first-order phase separation transition at finite $T$, terminating at a second-order critical point where $\chi_\text{charge}$ diverges. This critical point aligns with a quantum critical point ($T = 0$) at $t' = 0$, separating pseudogap and Fermi-liquid regions. Sordi et al.~\cite{sordi2012pseudogap} connect this behavior to the Widom line, a line of maxima in $\chi_\text{charge}$ that extends from the QCP, and which represents a thermodynamic crossover boundary, organizing the phase space and marking the onset of the pseudogap phase at a characteristic temperature $T^*$.

\vspace{0.25cm}
\noindent \textbf{Hole clustering in METTS snapshots} 

\noindent Simulations with iPEPS at finite temperature in the grand-canonical ensemble typically result in a uniform particle density across the system due to the choice of a translationally invariant ansatz. Studies done in the canonical ensemble using METTS~\cite{white2009, stoudenmire2010minimally} fix the total number of particles. As a result, charge inhomogeneities can readily be observed. Here, we follow the methods outlined in Ref.~\cite{wieteksquare2021}. We conducted our simulations primarily on cylinders of width $W=4,6$ (periodic) and length $L\leq 40$ (open). Using the same on-site Coulomb repulsion ($U = 10$), we set the electron filling to $n = 0.9375$. We chose this filling as it offered favorable convergence properties, while detailed studies in SI~\cite{SI} confirm that our conclusions are valid across a broad range of fillings. The METTS algorithm, a Markov-chain Monte Carlo method, simulates thermal states via iterative imaginary-time evolution of product states~\cite{white2009,stoudenmire2010minimally}. The evolution process involves creating METTS snapshots $|\psi_{i}\rangle$,
\begin{equation}
|\psi_i \rangle = \frac{1}{\sqrt{p_i}}e^{-\beta H/2} |\sigma_i\rangle, 
\label{thermal_evol}
\end{equation}
\noindent where $|\sigma_i\rangle$ denotes the basis of product states, $p_i=\langle \sigma_i | e^{-\beta H} | \sigma_i\rangle$, and $\beta=1/T$ the inverse temperature. By averaging over the individual observables calculated via snapshot wavefunctions $|\psi_{i}\rangle$, we estimate the thermal observables:
\begin{align}
\langle O \rangle &= \frac{1}{Z} \text{Tr}(e^{-\beta H} O) \nonumber\\
&= \frac{1}{Z} \sum_i  \langle \sigma_i | e^{-\beta H/2} O e^{-\beta H/2} | \sigma_i \rangle \nonumber\\
&=\frac{1}{Z} \sum_i p_i \langle \psi_i | O | \psi_i \rangle ,
\label{thermal_average}
\end{align}
where $Z$ is the partition function given by $Z = \sum_i p_i$.   Hole densities and spin correlations can be calculated directly from these snapshot wavefunctions. In representative snapshots in Fig.~\ref{fig:2}, we observe at panels (a)$T=0.200$, (b) $T=0.150$, and (c) $T=0.100$, large hole clusters and pronounced antiferromagnetic 
domains, indicative of strong hole clustering. At panel (d) $T=0.025$, we observe the stripe order with intertwined charge and spin density waves. Hole densities at a lattice site $r$ are calculated as $n_{h}(r) = 1 - \langle \psi_i | n_r | \psi_i \rangle$, where $n_r = n_{r\uparrow} + n_{r\downarrow}$ and the radius of the grey circle at a site $r$ is proportional to $n_{h}(r)$. Spin correlations are illustrated by red and blue arrows of length proportional to $\langle 
\psi_i | S^z_{0} \cdot S^z_{r} | \psi_i \rangle$, where the label $r=0$ is a reference site. Because SU(2) symmetry is intact, local moments in a thermal mixed state are small and basis dependent; we therefore visualize correlations rather than local $\langle S^z\rangle$. To represent staggered domains, the squares denoting lattice sites are color-coded; regions with staggered correlations are shaded in one color, with adjacent staggered domains differing by a $\pi$-phase shift shown in the alternate color (either red or blue). Across all panels we keep a fixed reference circle radius (hole density) and arrow length (spin correlator) to provide absolute visual scales. As thermal fluctuations diminish with decreasing temperature, the system increasingly favors the formation of more extensive antiferromagnetic domains (snapshots for wider cylinder of width $W=6$ are in the SI \cite{SI}). These pronounced fluctuations foreshadow the instability that ultimately condenses into stripe order. At the lower temperature $T=0.025$ (Fig.~\ref{fig:2}(d)), we no longer find stochastic clustering but instead a stripe order with wavelength $\sim 8$ lattice sites.  
\begin{figure*}[t!]
\vspace{-0cm}
\includegraphics[width=1.0\textwidth,clip=true]{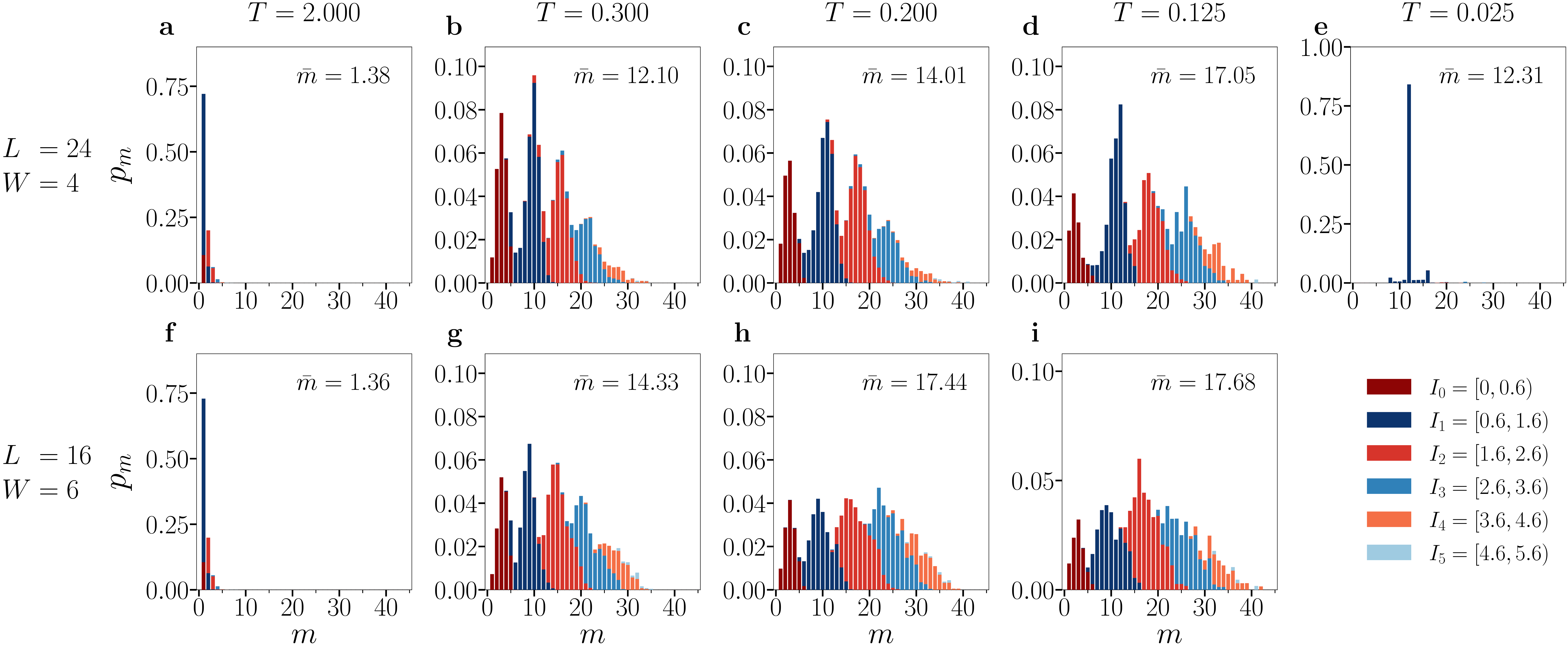}
\vspace{-0.2cm}
\caption{\textbf{Cluster statistics resolved by hole mass.}
Stacked histograms of the cluster–size distribution $p_m$ vs. size $m$. Colors encode the cluster hole mass $\rho(\mathcal C)=\sum_{r\in\mathcal C} n_h(r)$ binned into $\mathcal I=\{[0,0.6), [0.6,1.6), [1.6,2.6), [2.6,3.6),
[3.6,4.6), [4.6,5.6)\}$, so $p_m=\sum_{I\in\mathcal I} p_m^{(I)}$. Top row (a–e): Cylinder of size $24\times4$ at $T=2.000,\,0.300,\,0.200,\,0.125,\,0.025$. Bottom row (subplots f–i): Cylinder of size $16\times6$ at $T=2.000,\,0.300,\,0.200,\,0.125$. Both geometries have $N=96$ sites and are set to filling $n=0.9375$. At $T=2.0$ (subplots a and f), $p_m$ is nearly exponential, dominated by $m=1$ with $\rho(\mathcal C)\in[0.6,1.6)$. On cooling, weight shifts to larger $m$ and oscillatory lobes appear; lobe $k$ is mainly supplied by window $I_k$, consistent with AFM-assisted, near-integer hole aggregation. Because thresholding selects stripe cores ($\approx3$ sites within an $\approx 8$-site wavelength), a stripe segment yields $\rho(\mathcal C)\in[0.6,1.6)$ rather than $2$. The mean cluster size $\bar m$ is consistently larger for $W=6$ than $W=4$ at intermediate temperatures. At $T=0.025$ (subplot e), $p_m$ concentrates near cluster size $m=12$ (see Fig.~\ref{fig:2}(d) for a snapshot), indicating the onset of a stripe order.}
\label{fig:3}
\end{figure*}
To capture the clustering behavior, we employ an algorithm to statistically analyze the hole clusters in the METTS snapshots. We define the set of all lattice sites in a METTS snapshot $\ket{\psi_{i}}$ as $\Lambda$. Each snapshot is scrutinized to identify sites where the local hole density $n_h$ exceeds a specified threshold. This threshold $n_{h}^{\text{th}}$ is defined as 
\begin{equation}
 n_{h}^{\text{th}} = 1 - n + c\sigma_{n_h}. 
 \label{threshold}
\end{equation}
where $1-n$ is the mean hole density, $\sigma_{n_h}$ is the standard deviation of the hole density distribution within a METTS snapshot $\ket{\psi_{i}}$,
\begin{equation}
\sigma_{n_h} = \sqrt{\frac{1}{|\Lambda|} \sum_{r \in \Lambda} \left( n_h(r) - (1 - n) \right)^2},
\label{stddev}
\end{equation}
where $|\Lambda|$ is the total number of lattice sites. Here, $c$ is a sensitivity parameter set to $c=0.5$ to balance between detecting meaningful clusters and avoiding sensitivity to individual site variations. The sites satisfying this threshold condition are marked with black borders in METTS snapshots in Fig.~\ref{fig:2}. Our clustering analysis is robust to the threshold choice $c$ across $c \in [ 0.0, 0.9]$ (see SI~\cite{SI}), and while absolute cluster sizes vary, the temperature range where large clusters appear remains stable.

\vspace{0.2cm}
\noindent \textbf{Statistical analysis of clustering}

\noindent We analyze a single METTS snapshot by first identifying the set of sites
whose local hole density exceeds the threshold $n_h^{\text{th}}$,
\begin{equation}
\mathcal{E} = \{\,r\in\Lambda \mid n_h(r) > n_h^{\text{th}}\}\; .
\end{equation}
We partition $\mathcal E$ into disjoint clusters $\mathcal C$, defined as nearest-neighbor connected components of $\mathcal E$; two sites are connected if they share a horizontal or
vertical bond:
\begin{equation}
\mathcal{E} = \dot{\bigcup_{\mathcal C}} \,\mathcal C .
\label{clusters}
\end{equation}
For each cluster we record its size $m = |\mathcal C|$ and its total hole mass
\begin{equation}
\rho(\mathcal C)= \sum_{r\in\mathcal C} n_h(r).
\end{equation}

\noindent Collecting all clusters from all snapshots, the density–weighted cluster–size probability is defined as
\begin{equation}
p_{m} = \frac{
  \sum_{\substack{\text{snapshots}\\\mathcal C:\,|\mathcal C|=m}}
    \rho(\mathcal C)
}{
  \sum_{\substack{\text{snapshots}\\\mathcal C}}
    \rho(\mathcal C)
},
\qquad
\sum_{m}p_{m}=1.
\label{probability}
\end{equation}
\noindent We now ask which range of hole masses supply the weight at each cluster size $m$. To do so, we coarse-grain the cluster hole mass $\rho(\mathcal C)$ into non-overlapping windows defined by 
\begin{equation}
I_{0} = [0, s); \hspace{0.15cm} I_{k} = [k-1+s, k+s) \hspace{0.15cm} \forall  k = 1, 2, 3, \ldots,
\label{integer}
\end{equation}
and define $\mathcal I \equiv \{I_k\}_{k\ge 0}$. We then decompose
\begin{equation}
p_m=\sum_{I\in\mathcal I} p_m^{(I)},\hspace{0.1cm}
p_m^{(I)}=\frac{\displaystyle\sum_{\text{snapshots}}\sum_{\mathcal C:\,|\mathcal C|=m,\ \rho(\mathcal C)\in I}\rho(\mathcal C)}{\displaystyle\sum_{\text{snapshots}}\sum_{\mathcal C}\rho(\mathcal C)}.
\end{equation}
\noindent Figure~\ref{fig:3} shows stacked histograms of $p_m$ versus $m$, where the bar at each $m$ is built from the components $p_m^{(I)}$; thus the total height equals $p_m$ while the color encodes the prevailing $\rho(\mathcal C)$ window. The mean cluster size is
\begin{equation}
\bar m=\sum_m p_m\, m. 
\label{mcs}
\end{equation}
The top panel (subplots $a-e$) presents data from a cylinder of length $L=24$ and width $W=4$ at $T=2.000, 0.300, 0.200, 0.125, 0.025$. The bottom panel (subplots $f-i$) shows data from a cylinder of $L=16$ and $W=6$ at $T=2.000,\,0.300,\,0.200,\,0.125$ (no lower-$T$ data converged for this width). Both geometries have $N=96$ sites and are set at filling $n=0.9375$, hence the same total hole number $N_h=(1-n)N=6$. Therefore the differences across panels isolate finite-width effects at fixed area. With threshold 
$c=0.5$, we set $s=0.6$ in Eq. \ref{integer} because this choice best separates the lobes of $p_{m}$ seen in Fig.~\ref{fig:3}; adjacent lobes differ by roughly one hole of total mass. At $T=2.000$ the distribution is nearly exponential, dominated by $m=1$ with $\rho(\mathcal C) \in I_{1}=[0.6,1.6)$. Upon cooling below $T\approx 0.75$ (see SI~\cite{SI} for a full temperature sweep of clustering 
statistics), the weight shifts to larger $m$ and the histograms develop somewhat oscillatory, peaked lobes in $m$ (roughly $1-5$, $6-13$, $\ldots$). Successive lobes are predominantly supplied by successive mass windows in $\mathcal I$, i.e., moving from lobe $k$ 
to $k+1$ adds about one hole’s worth of charge to typical clusters. Physically, the antiferromagnetic background penalizes extended domain walls; aggregating carriers in near-integer increments lets holes assemble one–by–one while minimally disturbing local AFM order, producing the observed stepwise pattern. Because thresholding isolates the high–density cores of stripes (about three sites within an $\sim8$-site wavelength), a stripe segment contributes $
\rho(\mathcal C)\in I_{1} = [0.6,1.6)$ rather than $\approx 2$. The mean cluster size $\bar m$ is systematically larger for $W=6$ than for $W=4$, demonstrating that cluster sizes grow with width. Finite-size effects are thus separable: increasing $W$ at fixed area broadens $p_m$ and raises $\bar m$; increasing $L$ at fixed $W$ softens the large-$m$ cutoff and reveals additional lobes (see SI~\cite{SI}). Importantly, in the clustering window the probability that a cluster wraps around the cylinder width remains well below unity and decreases with $W$ (see SI~\cite{SI}). All of these are inconsistent with merely meandering stripe fluctuations at finite temperature. By contrast, below $T\approx 0.075$ (see SI~\cite{SI}) --- for example at $T=0.025$ as shown in subplot (e) for cylinder of size $L=24,W=4$ --- the distribution collapses near $m=12$, signaling entry into the stripe regime with a sharply selected cluster size.

\begin{figure}[!htbp]
\vspace{-0.0cm}
\includegraphics[width=1.0\columnwidth,clip=true]{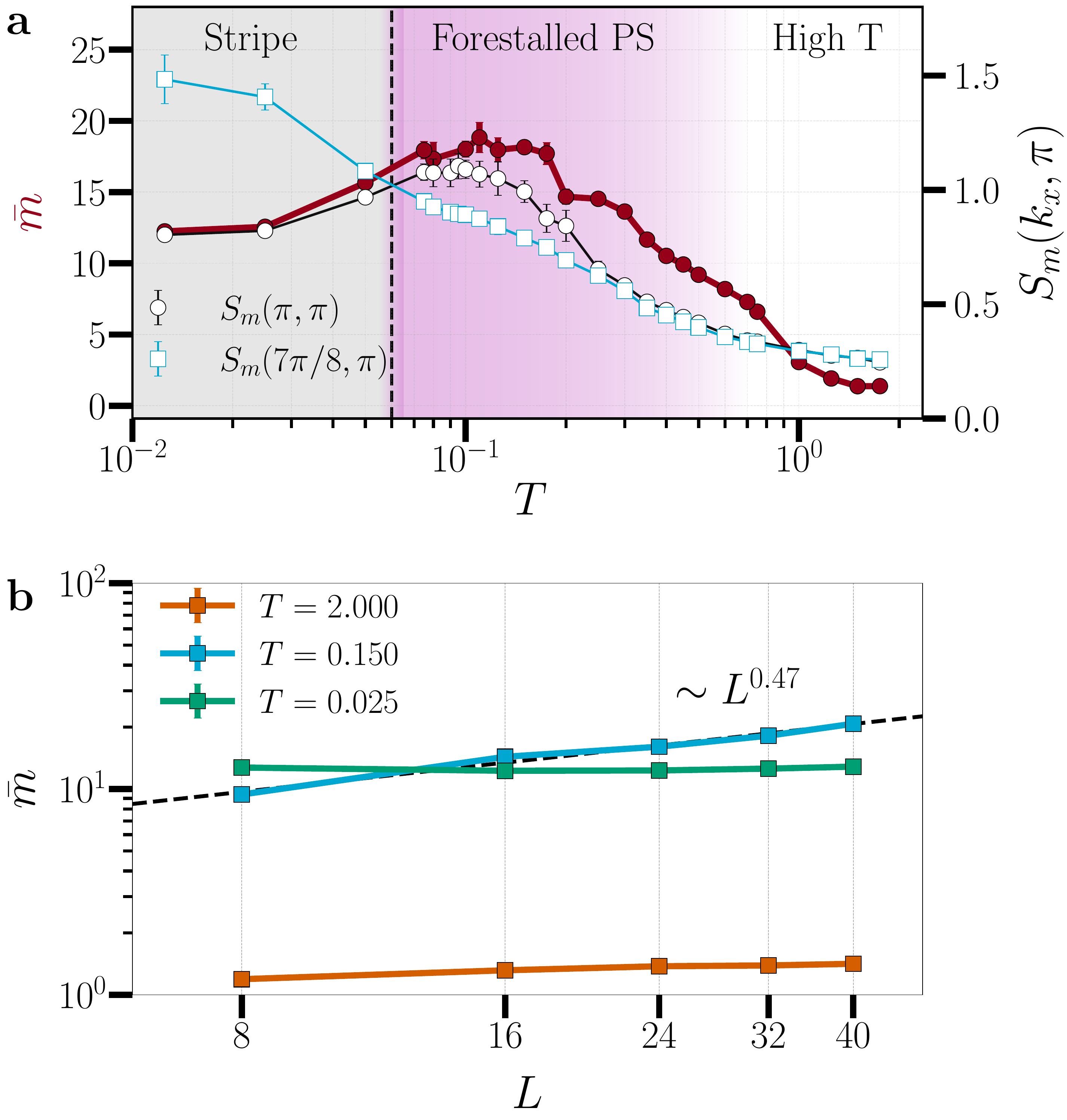}
\vspace{-0.3cm}
\caption{\textbf{Clustering and magnetic correlations in the finite–temperature Fermi–Hubbard model.}  
(a) Density–weighted mean cluster size $\bar m=\sum_m p_m m$ (maroon circles) and magnetic structure factors $S_{m}(\pi,\pi)$ (black open circles) and $S_{m}(7\pi/8,\pi)$ (blue open squares) versus temperature $T$ at $U=10$ and $n=0.9375$. The black dashed vertical line marks the temperature $T\approx 0.06$, where $S_{m}(7\pi/8,\pi)$ overtakes $S_{m}(\pi,\pi)$; we use this crossing as an operational definition of stripe onset. The shaded regions indicate the low–temperature stripe regime (grey) and the crossover window of forestalled phase separation (lavender), which smoothly fades into the high–temperature phase to the right. (b) Log–log scaling of $\bar m(L)$ across temperatures. In the clustering regime ($T=0.150$), $\bar m$ grows approximately as a power law with $L$, while at high temperature ($T=2.000$) and deep in the stripe phase ($T=0.025$) it remains nearly $L$–independent.}
\label{fig:4}
\end{figure}
\noindent The temperature dependence of the mean cluster size $\bar m$ is shown in Fig.~\ref{fig:4}(a) (additional dopings in SI~\cite{SI}). To connect charge clustering to magnetism, and to separate the clustering crossover from the onset of static stripes, we monitor the static magnetic structure factor,
\begin{equation}
S_m(\mathbf{k})=\frac{1}{N}\sum_{l,m=1}^{N}e^{i\mathbf{k}\cdot(\mathbf{r}_l-\mathbf{r}_m)}\langle S_l^z S_m^z\rangle,
\label{msf}
\end{equation}
\begin{figure}[!htbp]
\vspace{-0.2cm}
\includegraphics[width=0.99\columnwidth,clip=true]{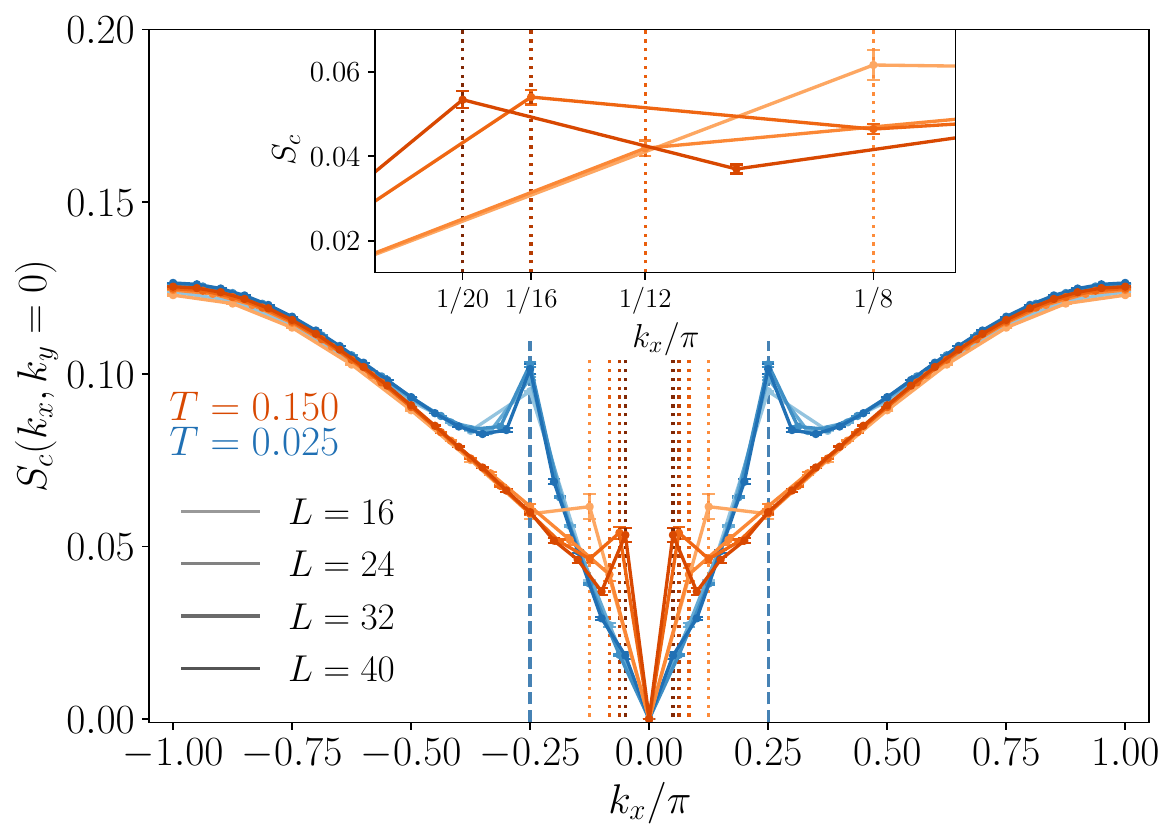}
\vspace{-0.4cm}
\caption{\textbf{Charge structure factor}. Static charge structure factor $S_{c}(\mathbf{k})$ (see Eq.~(\ref{csf})) for electron density $n = 0.9375$ and $U=10$ across system sizes of length $L = 16, 24, 32, 40$ and width $W = 4$. It manifests a subtle inner peak at $\mathbf{k} = (\frac{2\pi}{L},0)$ at temperature $T = 0.150$ resolved by the lattice length. It shifts inward with increasing $L$ (dotted orange lines). The inset highlights these shifts with darker shades for larger sizes. At $T = 0.025$, the onset of stripe order is marked by a stable peak at $\mathbf{k} = \left(\frac{\pi}{4}, 0\right)$ (blue dashed line) consistent across different system sizes.}
\label{fig:5}
\end{figure}
evaluated at two physically motivated momenta: (i) $(\pi,\pi)$ the Néel vector that tracks antiferromagnetic correlations; and (ii) $(7\pi/8,\pi)$, the stripe ordering vector observed when stripes form on $W=4$ cylinders at filling 
$n=0.9375$~\cite{wieteksquare2021}. Upon cooling from high $T$, $\bar m$ grows and develops a broad maximum within a crossover window (lavender) where $S_{m}(\pi,\pi)$ is also enhanced; this reflects the formation of extended AFM domains surrounding hole clusters, i.e.\ the forestalled phase separation regime. At lower $T$ the magnetic weight shifts from $(\pi,\pi)$ to the stripe vector: $S_{m}(7\pi/8,\pi)$ overtakes $S_{m}(\pi,\pi)$ (marked by a black dashed vertical line at $T\approx0.06$) as static stripe correlations set in (grey shading). Concomitantly, $\bar m$ stops broadening and locks to the characteristic stripe cluster size, indicating that fluctuating clusters give way to a regular stripe pattern. This $(\pi,\pi)\!\to\!(7\pi/8,\pi)$ shift with decreasing $T$ is consistent with the finite–$T$ evolution reported for the same geometry in Ref.~\cite{wieteksquare2021} and with the emergence of the charge peak at $(\pi/4,0)$ in Fig.~\ref{fig:5}.
Figure~\ref{fig:4}(b) shows finite–size scaling of $\bar m(L)$ for width $W=4$. In the clustering window ($T=0.150$) we find sublinear power–law growth $\bar m(L)\sim L^{\gamma}$ with $\gamma\approx 0.47$, indicative of broad, system–spanning fluctuations. By contrast, at higher ($T=2.000$) and very low temperatures ($T=0.025$), there is almost no size dependence. For the latter, $\bar{m}(L)$ is locked at a value of around $12$, characterizing the stripe cluster size.

\vspace{0.225cm}

\noindent \textbf{Charge structure factor}

\noindent Finally, we explore a more conventional approach to detecting phase separation. In Fig.~\ref{fig:5} we present the equal-time (static) charge structure factor, 
\bea
S_c(\mathbf{k}) = \frac{1}{N} \sum_{l,m=1}^{N} e^{i\mathbf{k} \cdot (\mathbf{r}_l - \mathbf{r}_m)} \langle (n_l - n)(n_m - n) \rangle
\label{csf}
\eea
for a filling of $n = 0.9375$, considering various system sizes of lengths $L = 16, 24, 32$, and $40$, with a fixed width $W = 4$. Note that, since we are working in the canonical ensemble, $S_{c} (\mathbf{k}=0)=0$, because there is no fluctuation in total density at zero momentum. At a temperature $T = 0.150$, an inner peak is present at the lowest momentum mode $\mathbf{k} = (\frac{2\pi}{L},0)$ resolved by the lattice, which shifts inward as the system size increases. Physically, this inward shift with system size indicates a build-up of long-wavelength density fluctuations, characteristic of PS. This trend is highlighted by dotted orange lines that mark peak positions in both the main plot and inset. The inset further amplifies the inner peaks corresponding to different system sizes, using darker shades to represent larger sizes. Because the canonical ensemble fixes the total density, $S_c(\mathbf{k}=0)$ must vanish, which may impose an artificial constraint at the lowest nonzero momentum, i.e. $\mathbf{k} = (2\pi/L, 0)$. However, benchmarking against free fermions and varying $U$ shows that the marked low-$k$ peak emerges only at large $U$, indicating that it reflects genuine interaction-driven density fluctuations rather than an ensemble artifact (see SI~\cite{SI}).
  As the temperature decreases to $T = 0.025$, the system transitions into a stripe order, characterized by a pronounced peak at $\mathbf{k} = (\frac{\pi}{4}, 0)$ (illustrated by a blue dashed line) that remains consistent across different system sizes. \\
\vspace{0.25cm}

\noindent\textbf{Discussion}

\noindent We have used tensor-network simulations to map the finite-temperature phase diagram of the two-dimensional Fermi–Hubbard model in the pseudogap and strange-metal regimes.  At intermediate temperatures we find pronounced hole clustering coexisting with robust antiferromagnetic (AFM) domains, signaling a strong tendency toward phase separation.  Yet, upon further cooling, these clusters never coalesce into a macroscopic phase-separated state; instead, stripe order emerges.  In the canonical ensemble this “forestalled” phase separation appears as a low-momentum peak in the charge-structure factor.

AFM correlations have long been connected to pseudogap physics~\cite{Gunnarson2015,Schaefer2021,wieteksquare2021,xiao2023,meixner2024,simkovic2024}.  
Using self-consistent constrained-path AFQMC on large periodic lattices, Xiao et al.~\cite{xiao2023} showed that the doped Hubbard model evolves from a high-$T$ disordered metallic state, nearly uniform charge and only short-range AFM fluctuations, through growing commensurate AFM correlations, into incommensurate spin-density waves, and finally a finite-$T$ stripe order where charge order is locked to the spin pattern. Our results establish an intermediate-$T$ \textit{clustering} window between the disordered metal and static stripes: carriers aggregate into large, fluctuating hole-rich domains with enhanced $S_{m}(\pi,\pi)$ and a thermodynamic maximum in $\chi_{\text{charge}}$, yet macroscopic PS is preempted by stripe order.

Antiferromagnetic correlations are strongest when two electrons occupy neighboring sites; consequently, they favor half-filled regions and generate an effective attraction that empties neighboring areas, producing hole-rich clusters. Consistent with this, our hole-mass–resolved cluster-size histograms show stepwise aggregation: weight shifts through distinct lobes as clusters grow in near-integer (single-hole) increments at intermediate temperatures. On further cooling, the system selects a stripe order characterized by antiphase domain walls that favor pairwise rather than single-hole additions; so charge patterns settle without developing macroscopic phase separation. Correspondingly, the density-weighted mean cluster size increases in the clustering window and then settles to a selected value at low temperature. Resonant inelastic X-ray scattering (RIXS) experiments reveal high-temperature precursor charge density wave fluctuations and stripe correlations in cuprates~\cite{miao2017high,miao2018,miao2019high,arpaia2021charge}, lending support to the idea that charge fluctuations play an active role in the pseudogap regime.  Within AFM domains the strong spin background opens a gap, while adjacent hole-rich metallic regions introduce low-energy states that partially fill it, reducing the density of states at the Fermi level—an archetypal pseudogap hallmark.  While a detailed exploration of this mechanism is beyond our present scope, our results suggest that the interplay of AFM order and hole clustering may influence pseudogap formation.

In the $t$–$J$ model, when $J/t\!\ge\!1$, the ground state phase-separates~\cite{rommer2000,shih2001}.  Recent work shows that, under long-time evolution, charge degrees of freedom then remain essentially frozen~\cite{Luke2024}.  This immobility may be related to the large resistivity observed in cuprates across the pseudogap, strange-metal, and bad-metal regimes~\cite{kokalj2017,vuvcivcevic2023charge}.

\section*{Methods}

\label{methods}
 
\noindent Solving the FHM presents substantial challenges due to its strongly interacting nature; analytical solutions are rare and limited to specific cases~\cite{elliot1968,nagaoka}. Even the most advanced 
numerical methods often face difficulties in accurately solving the model without biases from approximation schemes~\cite{qin2022hubbard}. For finite temperature, the minimally entangled typical thermal states (METTS) algorithm, when applied to MPS~\cite{white2009,stoudenmire2010minimally}, offers significant advantages over traditional methods (such as purification~\cite{verstraete2004matrix}), notably by representing thermal states with lower bond dimensions. METTS has facilitated insightful studies into the thermal phase 
diagram of the square-lattice FHM including the study of the pseudogap regime and the stripe 
phase~\cite{wieteksquare2021}. Similarly, purification techniques using 
iPEPS~\cite{czarnik2012,czarnik2014,kshetrimayum2019, czarnik2021} have explored the phase diagram from 
high to intermediate temperatures, revealing significant distortions in antiferromagnetism upon 
doping~\cite{sinha2022}. Additionally, alternative approaches for finite temperature simulations using 
tensor networks for the Hubbard model include the exponential thermal tensor renormalization group 
(XTRG)~\cite{xtrg2021quantum}, tangent space tensor renormalization group (tanTRG)~\cite{tanTRG2023}, and 
METTS applied to PEPS~\cite{sinha2024}. A recent study using diagrammatic Monte Carlo~\cite{fedor2022} was 
able to achieve temperatures as low as $T=0.067$ and up to $U=7$ for arbitrary large lattices, providing 
insights into momentum-resolved spin and charge susceptibilities. See Ref.~\cite{leblanc2015} for a 
detailed list of methods approaching the FHM at finite temperatures. Complementing these computational breakthroughs, experimental techniques with ultracold atoms have similarly advanced, enabling precise 
simulation and investigation of many-body physics that echo the complex interactions found in the Hubbard model~\cite{esslinger2010fermi,BOHRDT2021168651,bakr2009quantum, sherson2010single, gross2021quantum,cheuk2015, parsons2015}. These experiments confine systems with hundreds of fermions and can reach temperatures as low as $1/4$ of the hopping energy, hosting non-trivial charge and spin correlations~\cite{boll2016spin, parsons2016site,hilker2017revealing, koepsell2019imaging, chiu2019string, koepsell2021microscopic}. 

\noindent In this work, we utilize two tensor network (TN) methods to simulate the FHM, focusing on approaches that are constrained by entanglement entropy, typically characterized by the bond dimension $D$. 

\vspace{0.2cm}
\noindent \textbf{Purification with Infinite Projected Entangled Pair States} \\
\noindent The charge susceptibility calculations in Fig.~\ref{fig:1} 
of the main text were obtained with the infinite projected entangled pair states (iPEPS) ansatz, using the neighborhood tensor 
update (NTU) algorithm~\cite{verstraete2011, Lubasch2014, dziarmaga2021}, as described in 
Ref.~\cite{sinha2022}. The iPEPS operates in the thermodynamic limit, effectively eliminating finite-size 
effects. We focus on local updates for iPEPS optimization, called the neighborhood tensor update 
(NTU)~\cite{dziarmaga2021}. This method is computationally more efficient and numerically stable compared 
to global updates like the Full Update. Additionally, NTU provides greater accuracy than the Simple 
Update~\cite{orus2014practical} which does mean-field-like approximations. The iPEPS approach we used 
simulates thermal states using the purification technique, which is broadly favored for finite-temperature simulations. In iPEPS, the purification of a thermal state is performed using a tensor network in which the thermal density matrix $\rho(\beta)$, representing the system at inverse temperature $\beta = 1/T$, is encoded as a pure state in an enlarged Hilbert space that includes both physical and ancillary degrees of freedom (d.o.f.). The process begins by representing the infinite temperature state as a product of maximally entangled states between corresponding local physical and ancillary d.o.f. This state serves as the starting point for imaginary-time evolution, implemented through a sequence of tensor network operations that progressively cool the system to the desired temperature. The thermal density matrix is then obtained by tracing out the ancillary degrees of freedom from the pure state representation. Mathematically, this is described by the equation:
\[
\rho(\beta) = \operatorname{Tr}_{\text{anc}}\left(|\psi(\beta)\rangle \langle \psi(\beta)|\right),
\]
where $|\psi(\beta)\rangle$ is the state obtained after applying the imaginary-time evolution operator $e^{-\beta H/2}$ to the initial state, and $\operatorname{Tr}_{\text{anc}}$ denotes the partial trace over the ancillary degrees of freedom. In our simulations, we enforced $U(1) \times U(1)$ symmetry (which conserves particle number) using the \texttt{YASTN} package~\cite{10.21468/SciPostPhysCodeb.52, 10.21468/SciPostPhysCodeb.52-r1.2}, optimized for fermionic and symmetric PEPS. This symmetry conservation helps to reduce computational overhead. With this approach, we could go up to bond dimensions $D = 30$ and were able to explore temperatures as low as $T = 0.125$. However, the rapid growth of the bond dimension due to the entanglement introduced by purification limits iPEPS's ability to access very low temperatures without significant computational cost.
\vspace{0.2cm}

\noindent \textbf{Minimally Entangled Typical Thermal States with Matrix Product States} \\
\noindent The second method we employed is the minimally entangled typical thermal states (METTS) algorithm, which uses matrix product states (MPS) as the variational ansatz. METTS is especially effective in 1D cylindrical geometries and excels at lower temperatures. Unlike purification, METTS does not require the full 
representation of the thermal density matrix. Instead, METTS relies on generating a sequence of typical thermal states through a Monte Carlo sampling process. The method works by selecting a product state, evolving it in imaginary time by $\beta/2$, and then collapsing the evolved state into a new 
product state using Markov chain sampling (see Eq. $(2)$ and Eq. $(3)$ of the main text). The imaginary time evolution in METTS is carried out using the time-dependent variational 
principle (TDVP)~\cite{haegeman2011, PAECKEL2019167998}, with subspace expansion~\cite{yang2020}, which improves the stability and accuracy of the evolution process. By avoiding the full tracking of the 
thermal density matrix, METTS can achieve accurate results with significantly lower bond dimensions than 
purification. Our METTS implementation, based on the 
\texttt{ITensor} library~\cite{10.21468/SciPostPhysCodeb.4-r0.3,10.21468/SciPostPhysCodeb.4}, was able to reach bond dimensions up to $D = 2500$, particularly for systems on $32 \times 4$ cylinder geometries, following successful approaches from previous studies~\cite{wieteksquare2021, wietektriangular2021}. A further advantage of METTS is its ability to provide spatially resolved data, which was key in identifying the clustering at intermediate temperatures.

\section*{Code availability}
The iPEPS simulations were performed with the YASTN library~\cite{10.21468/SciPostPhysCodeb.52, 10.21468/SciPostPhysCodeb.52-r1.2}. The MPS-METTS calculations used \texttt{METTS.jl} (\url{https://github.com/awietek/METTS.jl}), built on top of the ITensor library~\cite{10.21468/SciPostPhysCodeb.4-r0.3,10.21468/SciPostPhysCodeb.4}. We provide a minimal script for the Fermi–Hubbard cylinder and a README that reproduces the main figures.

\begin{acknowledgments}
We thank Antoine Georges, George Batrouni, Salvatore Manmana, Roderich Moessner, Steve White, Luke Staszewski, Martin Ulaga, Arnab Das, Joe H. Winter, Timon Hilker, Lode Pollet, Zhenjiu Wang, Thomas Chalopin, Immanuel Bloch, Fabian Grusdt, and Annabelle Bohrdt for insightful discussions and in particular Fakher Assaad for highlighting the signatures of phase separation in the charge structure factor. We thank Chris Hooley for suggesting the adjective ``forestalled'' and further discussions. A.W.\ acknowledges support by the DFG through the Emmy Noether program (Grant No.\ 509755282). A.S. thanks Marek M. Rams and Jacek Dziarmaga for previous technical discussions on iPEPS. A.S. acknowledges the Alexander von Humboldt Foundation for support under the Humboldt Research Fellowship and support from National Science Centre (NCN), Poland under project 2019/35/B/ST3/01028, at Jagiellonian University, Kraków, where part of the work was completed.
\end{acknowledgments}

\clearpage
\onecolumngrid            
\appendix                  
\section*{Supplementary Material}
\addcontentsline{toc}{section}{Supplementary Material} 

\setcounter{section}{0}
\setcounter{figure}{0}
\setcounter{table}{0}
\setcounter{equation}{0}
\renewcommand{\thesection}{S\arabic{section}}
\renewcommand{\thefigure}{S\arabic{figure}}
\renewcommand{\thetable}{S\arabic{table}}
\renewcommand{\theequation}{S\arabic{equation}}

\def\be{\begin{equation}}
\def\ee{\end{equation}}
\def\bea{\begin{eqnarray}}
\def\eea{\end{eqnarray}}
\def\bi{\begin{itemize}}
\def\ei{\end{itemize}}

\title{Supplementary Information for ``Forestalled Phase Separation as the Precursor to Stripe Order''}

\author{Aritra Sinha} 
\affiliation{Max Planck Institute for the Physics of Complex Systems, 
             N\"{o}thnitzer Strasse 38, Dresden 01187, Germany}
             
\author{Alexander Wietek} 
\affiliation{Max Planck Institute for the Physics of Complex Systems, 
             N\"{o}thnitzer Strasse 38, Dresden 01187, Germany}

\date{\today}

\maketitle

\begin{figure*}[!htbp]
\vspace{-0cm}
\includegraphics[width=0.98\textwidth,clip=true]{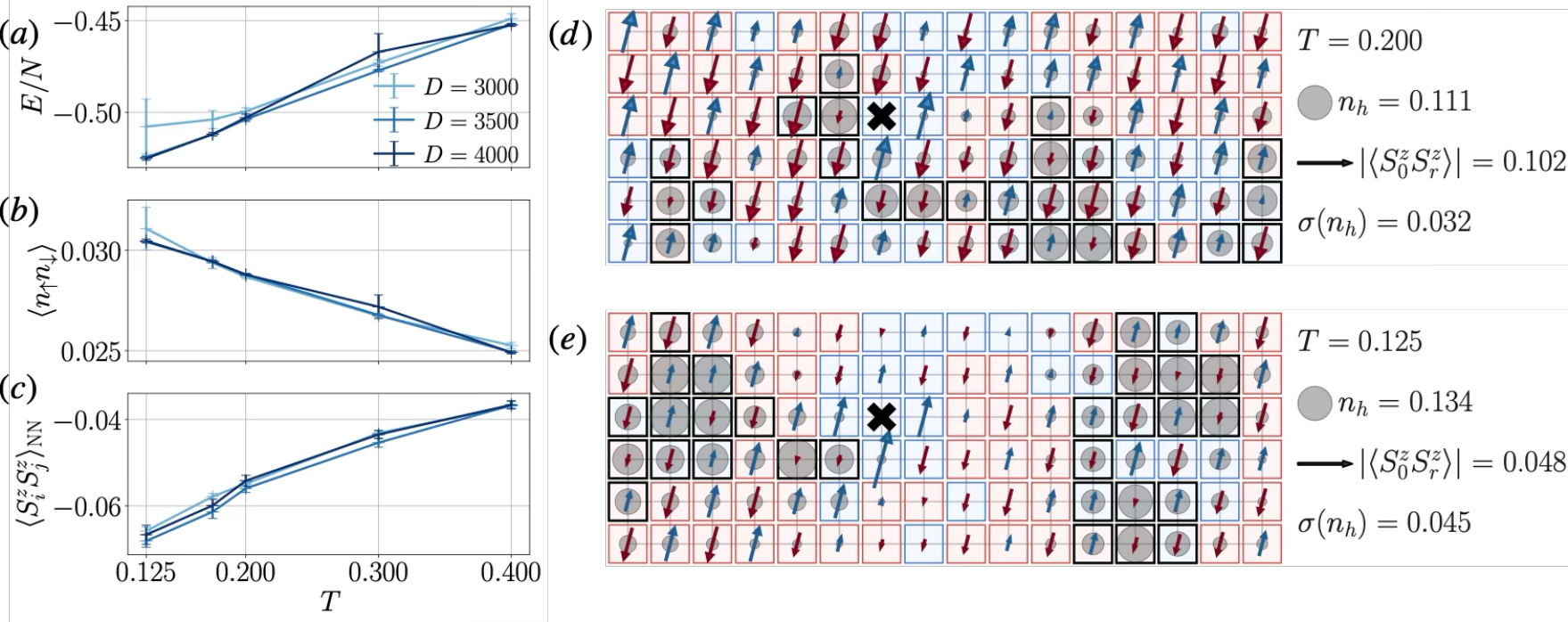}
\vspace{-0cm}
\caption{\textbf{Convergence of larger width cylinder and METTS snapshots}. (a) Energy per site $E/N$, (b) double occupancy $\langle n_{\uparrow}n_{\downarrow}\rangle$ and (c) nearest-neighbor spin-spin correlation $\langle S_{i}^{z} S_{j}^{z} \rangle_{\text{NN}}$ as functions of bond dimension for a cylinder of length $L=16$ and width $W=6$ at electron density $n=0.9375$ and on-site repulsion $U=10$ at various temperatures. The data is well converged up to $T=0.125$. Typical METTS snapshots for same parameters at temperatures (d) $T=0.2$ and (e) $T=0.125$ showing hole clustering and antiferromagnetic domains, similar to observations on narrower cylinders.}
\label{fig:S3}
\end{figure*}

\noindent In this Supplementary Information we compile seven sections of additional results that support or extend results of the main text. Section~\ref{I} benchmarks our finite-temperature iPEPS calculations by demonstrating bond-dimension convergence (Fig.~\ref{fig:S1}) and tracking the temperature evolution of the peak in the charge susceptibility (Fig.~\ref{fig:S2}). In Section~\ref{II}, we focus on the METTS analysis of wider cylinders, documenting convergence and representative snapshots on a $16\times6$ lattice (Fig.~\ref{fig:S3}). A dedicated Section~\ref{III} presents the complete temperature sweep of density-weighted cluster-size histograms on a $32\times4$ cylinder, which also makes the low-temperature stripe formation explicit (Fig.~\ref{fig:S4}). Section~\ref{IV} probes the robustness of the cluster definition; Fig.~\ref{fig:S5} summarizes how the mean cluster size depends on the threshold parameter $c$ and on the filling. Section~\ref{V} disentangles ensemble and interaction effects in the charge structure factor via side-by-side grand-canonical analytics, canonical free-fermion snapshots, and interacting ($U=10$) METTS (Fig.~\ref{fig:S6}), together with the interaction-strength dependence at fixed $T$ (Fig.~\ref{fig:S7}). Section~\ref{VI} examines next-nearest-neighbor hopping: representative METTS snapshots and cluster statistics for $t'=0.3$ are shown in Figs.~\ref{fig:S8} and \ref{fig:S9}. Finally in Section~\ref{VII} we further address the finite-size effects of our METTS results by quantifying cylinder width wrap-around probabilities of clusters and the cylinder length dependence of cluster statistics (Figs.~\ref{fig:S10} and \ref{fig:S11}).

\section{iPEPS Convergence and Temperature Dependence of Charge Susceptibility}
\label{I}

\begin{figure}[!htb]
\vspace{-0cm}
\includegraphics[width=0.6\columnwidth,clip=true]{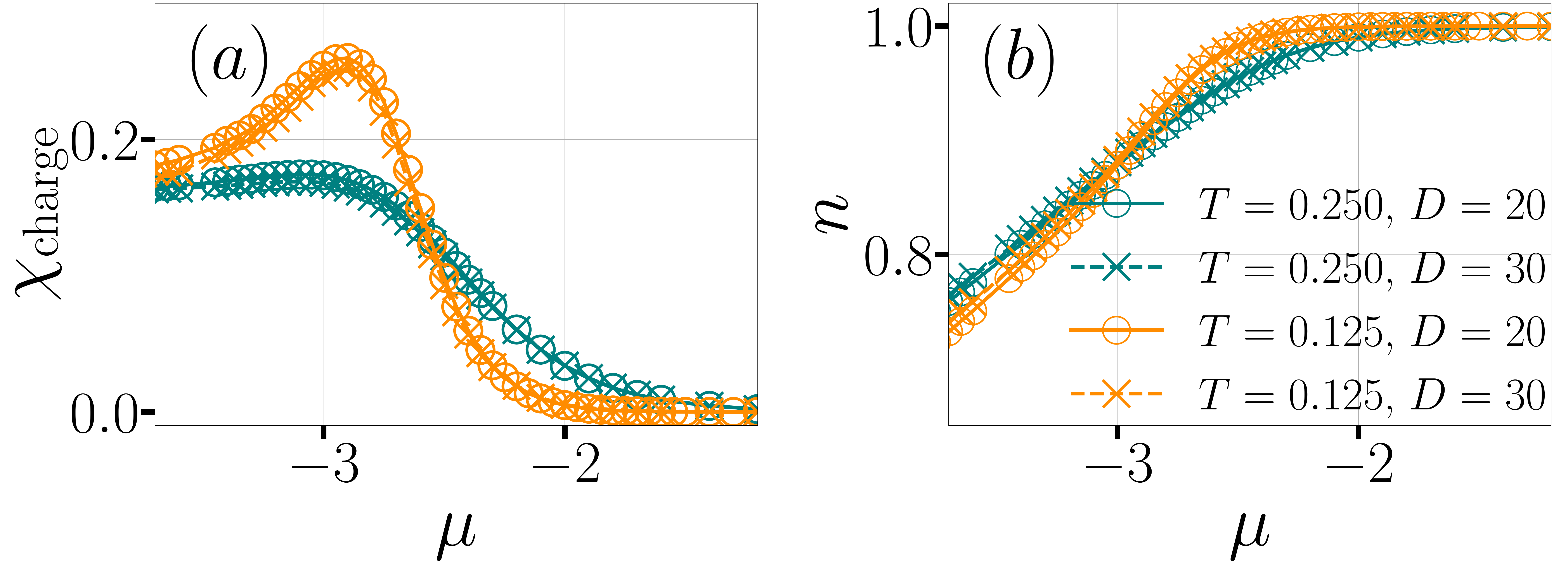}
\vspace{-0.4cm}
\caption{\textbf{Convergence of Charge Susceptibility and Density with Bond Dimension.} (a) Charge susceptibility, $\chi_{\mathrm{charge}} = \frac{dn}{d\mu}$, and (b) filling $n$ as functions of chemical potential $\mu$ for bond dimensions $D=20$ (circles) and $D=30$ (crosses), at temperatures $T = 0.25$ (teal) and $T = 0.125$ (dark orange). This demonstrates that features highlighted in main text Fig.~$1$ such as the peak in susceptibility have converged.
}
\label{fig:S1}
\end{figure}
In the main text Fig.~$1$, we perform simulations using the purification~\cite{verstraete2004matrix} technique with the iPEPS ansatz to calculate the charge susceptibility $\chi_{\mathrm{charge}} = \frac{\partial n}{\partial \mu}$ and electron density $n$ as functions of chemical potential $\mu$ at various temperatures. We showed that a maximum in $\chi_{\mathrm{charge}}$ emerges starting from intermediate temperatures, indicating enhanced susceptibility towards potential phase separation. These calculations were performed with bond dimension $D=20$. Fig.~\ref{fig:S1} shows both the $\chi_{\mathrm{charge}}$ and $n$ for bond dimensions $D=20$ and $D=30$ at temperatures $T=0.25$ and $T=0.125$. The close correspondence between the results obtained at $D=20$ and $D=30$ indicates that our calculations are reliably converged with respect to bond dimension, affirming the robustness of the features observed in $\chi_{\mathrm{charge}}$.

\begin{figure}[!htb]
\vspace{-0cm}
\includegraphics[width=0.6\columnwidth,clip=true]{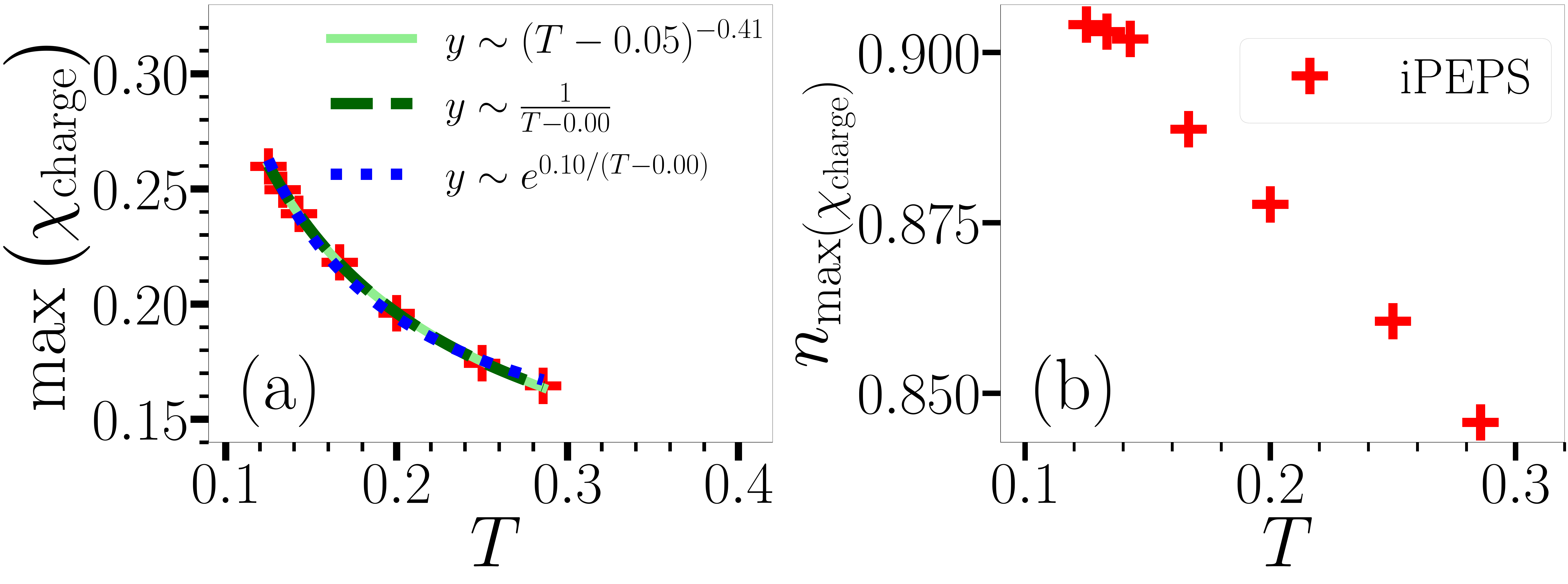}
\vspace{-0cm}
\caption{\textbf{Analysis of Peak Charge Susceptibility.} (a) The peak value in charge susceptibility $\chi_\text{charge}$ as a function of temperature. Power-law fit with $(T-T')^{\gamma}$ (solid green line) and linear fit with $1/(T-T')$ (dashed dark green line) capture a rapid growth and do not exclude a divergence of $\chi_\text{charge}$ as $T \rightarrow T_{c} \geq 0$. (b) Filling corresponding to the maximum charge susceptibility, $n_{\max(\chi_{\text{charge}})}$ as a function of temperature $T$.
}
\label{fig:S2}
\end{figure}

\noindent Fig.~\ref{fig:S2}(a) displays the peak charge susceptibility $\max{(\chi_{\text{charge}})}$ as a function of temperature. Upon cooling, $\max{(\chi_{\text{charge}})}$ rises steadily; the trend can be fitted equally well by a linear form $1/(T-T')$ and by a power-law $(T-T')^{\gamma}$. The good agreement of both fits with the numerical points implies that $\chi_{\text{charge}}$ may diverge as $T$ approaches a critical value $T_c \ge 0$, signaling a strong enhancement of charge fluctuations that we attribute to incipient hole clustering. Such clustering serves as a precursor to phase separation; however, as discussed in the main text, the onset of stripe order at lower temperatures ultimately prevents full phase separation. In Fig.~\ref{fig:S2}(b), we plot the density corresponding to the maximum of charge susceptibility, $n_{\max({\chi_\text{charge}})}$ as a function of temperature. 

\section{Convergence for wider METTS Cylinders}
\label{II}

In the main text, we used METTS simulations to study the formation of hole clusters at intermediate temperatures ($0.5 \lesssim T \lesssim 0.075$) and the emergence of stripe order at very low temperatures ($T \lesssim 0.05$) on cylinders of width $W=4$. To check whether these features persist at larger widths, we perform additional METTS simulations on a $16\times 6$ cylinder. Converging a width‐$6$ system with matrix product states (MPS) is more challenging because the entanglement entropy $S$ across a bipartition scales as $S \propto W$ in two‐dimensional systems represented by one‐dimensional MPS~\cite{eisert2010colloquium}. In practice, $S\approx \alpha W$, so the required bond dimension grows like $D\propto e^{S}\approx e^{\alpha W}$~\cite{schollwock2011density}.

Fig.~\ref{fig:S3} shows (a) the energy per site, (b) the average double occupancy and (c) nearest-neighbor spin-spin correlation $\langle S_{i}^{z} S_{j}^{z} \rangle_{\text{NN}}$ versus temperature for the $16\times 6$ cylinder at electron density $n=0.9375$ and on-site repulsion $U=10$ using bond dimensions $D=3000,3500,4000$. The close overlap down to $T=0.125$ indicates convergence. In Fig.~\ref{fig:S3}(d) and (e), we have representative METTS snapshots for $T=0.2$ and $T=0.125$. It shows strong hole clustering and large antiferromagnetic domains. Black‐bordered sites mark hole densities above the threshold in Eq.~$4$ of the main text (or Eq.~\ref{th} here with $c=0.5$), and adjacent black‐bordered sites are grouped into nearest–neighbor clusters $\mathcal C$ [Eq. $7$ in the main text]; for each cluster we record its size $m=|\mathcal C|$ and hole mass $\rho(\mathcal C)=\sum_{r\in\mathcal C} n_h(r)$. We recall the density–weighted size distribution (Eq. $10$ and Eq. $11$ in the main text)
\[
p_m=\frac{\sum_{\text{snapshots}}\sum_{\mathcal C:\,|\mathcal C|=m}\rho(\mathcal C)}{\sum_{\text{snapshots}}\sum_{\mathcal C}\rho(\mathcal C)}\,,\qquad \sum_m p_m=1,
\]
its hole–mass decomposition $p_m=\sum_{I\in\mathcal I}p_m^{(I)}$.

\section{Clustering statistics for a wide range of temperatures}
\label{III}

\begin{figure}[!htbp]
  \centering
  \includegraphics[width=0.9\columnwidth]{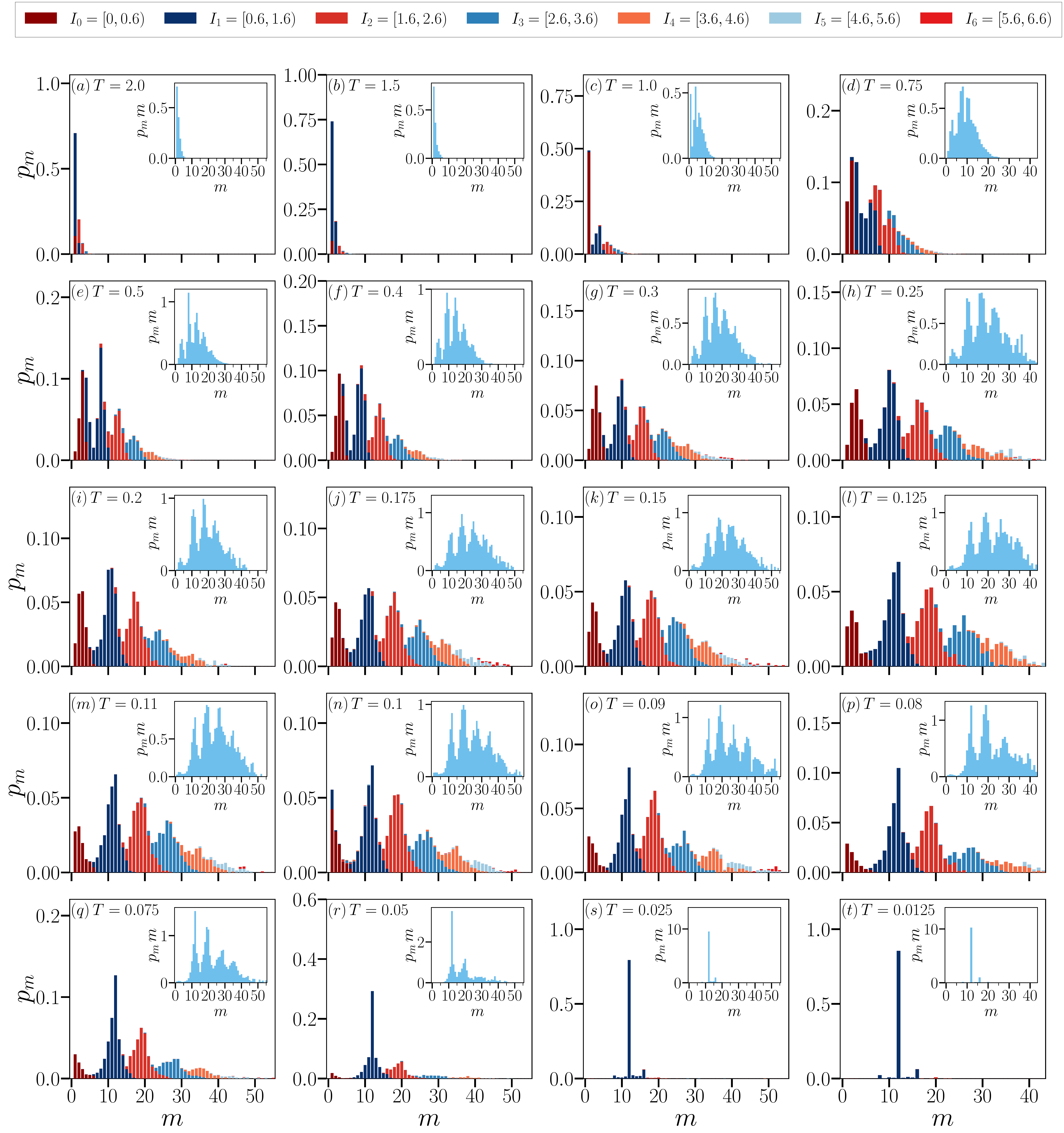}
  \caption{\textbf{Evolution of cluster‐size distributions in METTS snapshots across temperatures.}  
    Stacked histograms of the density–weighted cluster–size distribution $p_m$ vs. size $m$ for a cylinder of size $32\times4$ at $U=10$, $n=0.9375$, with temperatures labeled in each subplot. Colors encode the cluster hole mass $\rho(\mathcal C)=\sum_{r\in\mathcal C} n_h(r)$ binned into $\mathcal I=\{[0,0.6),[0.6,1.6),[1.6,2.6),[2.6,3.6),[3.6,4.6),[4.6,5.6)\}$, so $p_m=\sum_{I\in\mathcal I} p_m^{(I)}$. Insets show $p_m m$ to highlight the total contribution of each cluster. This comprehensive temperature evolution plot broadly distinguishes between the three characteristic regimes discussed in the main text: disordered (high T), cluster-rich (intermediate T), and stripes (low T). At high temperatures ($T=2.0,1.5, 1.0$) large clusters are exponentially suppressed;  
    at intermediate temperatures ($T=0.5$-$0.075$) the distributions broaden; and at the lowest temperatures ($T=0.05,0.025,0.0125$) a pronounced peak at $m=12$ emerges, marking the onset of charge density wave clusters.}
  \label{fig:S4}
\end{figure}

As a comprehensive analysis of how hole clustering evolves across the entire temperature range studied, Fig.~\ref{fig:S4} systematically presents the temperature-dependent evolution of the stacked histograms of density-weighted cluster-size distributions, $p_m$ vs cluster size $m$ (described in details in the main text), obtained from METTS simulations on a $32\times4$ cylinder at strong coupling ($U=10$) and a fixed doping of $n=0.9375$. At the highest temperatures ($T \geq 1.0$), hole clusters are small and their sizes are exponentially suppressed, reflecting largely uncorrelated holes. Upon reducing the temperature into the intermediate regime ($T=0.5$ down to $T=0.075$), the distributions distinctly broaden, signaling the emergence of extended hole clusters. Finally, at the lowest temperatures ($T\leq0.05$), the distribution becomes sharply peaked around a characteristic cluster size of $m=12$, marking the onset of a well-defined charge density wave (stripe-ordered) phase.

\section{Sensitivity Analysis of Cluster Definition and Robustness Across Doping Levels}
\label{IV}

\noindent In the main text, we defined hole clusters in the METTS snapshots based on a threshold hole density 
\bea
n_h^{\text{th}} = 1 - n + c \sigma_{n_{h}},
\label{th}
\eea
where $n$ is the average electron density, $\sigma_{n_{h}}$ is the standard deviation (Eq. $5$ of the main text) of the hole density distribution in a snapshot, and $c$ is a coefficient set to $0.5$. We analyzed the cluster sizes and their temperature dependence to understand the formation of hole clusters and their relation to magnetic correlations. 

\begin{figure}[t!]
\vspace{-0cm}
\includegraphics[width=0.7\columnwidth,clip=true]{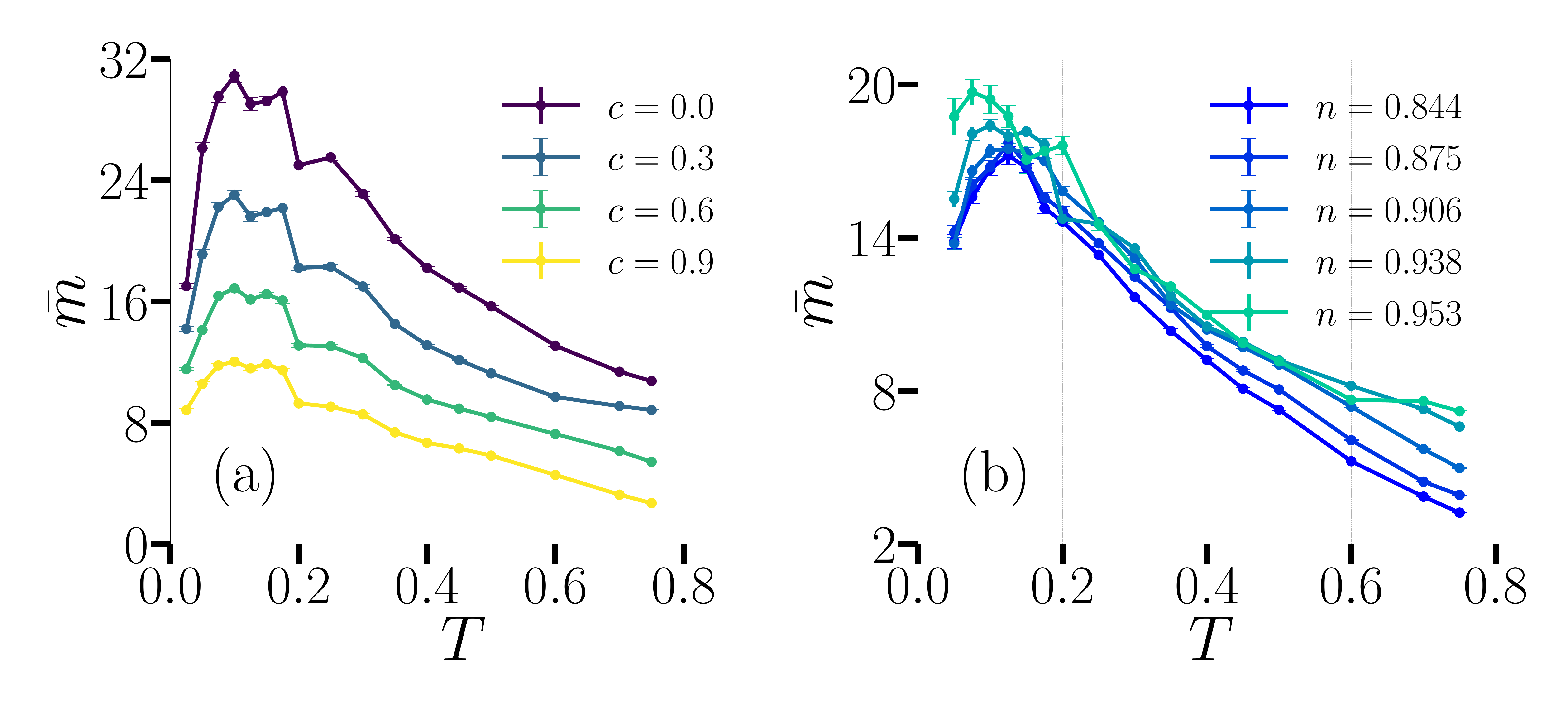}
\vspace{-0cm}
\caption{\textbf{Robustness of hole-clusters.} Cylinder of length $L=32$ and width $W=4$ at on-site repulsion $U=10$. (a) Mean hole cluster size $\bar{m}$ as a function of temperature $T$ at filling $n =0.9375$ for various values of the coefficient $c$ in the threshold $n_h^{\text{th}} = 1 - n + c \sigma_{n_{h}}$, which defines a hole cluster. (b) Clustering behavior at intermediate temperatures is observed over a range of densities from $n = 0.84375$ to $n = 0.95312$ and $c=0.5$.}
\label{fig:S5}
\end{figure}

To ensure that our results are not sensitive to the specific choice of $c$, we perform a robustness analysis by varying $c$ and examining the impact on the mean cluster size $\bar{m}$ (Eq. $9$ of the main text). Fig.~\ref{fig:S5}(a) shows the mean cluster size as a function of temperature for different values of $c$. The general behavior remains consistent across different values of $c$, with small clusters dominating at higher temperatures and larger clusters becoming prevalent at lower temperatures.

Fig.~\ref{fig:S5}(b) demonstrates that the maximum clustering behavior persists across a broad range of electron densities, specifically from $n = 0.84375$ to $n = 0.95312$. We have used bond dimension $D=2000$ throughout. This doping window covers both the strange metal and pseudogap regimes, indicating that charge clustering is a robust feature at finite temperature within these experimentally relevant phases of the Fermi-Hubbard model.

\section{Charge Structure Factor: Ensemble and Interaction Dependencies}
\label{V}

\noindent In the main text (Fig.~$5$) we defined the charge structure factor
\begin{equation}
S_c(\mathbf{k}) = \frac{1}{N} \sum_{l,m} e^{i \mathbf{k} \cdot (\mathbf{r}_l - \mathbf{r}_m)} \langle (n_l - n)(n_m - n) \rangle,
\label{csf}
\end{equation}
which exhibits an inner peak at $\mathbf{k}=(2\pi/L,0)$ for METTS simulations done in the canonical ensemble at strong coupling. In the canonical ensemble 
the total density is fixed, so by construction 
$S_c(\mathbf{k}=0)=0$.  If the inner peak at the smallest nonzero momentum were purely an artifact of enforcing $S_c(0)=0$, one would observe a small but 
broad redistribution of weight toward $k_x=2\pi/L$ even in the non-interacting limit.  

To understand this, we compare the $U=10$ METTS results with canonical and grand-canonical results for the free-fermion case ($U=0$) for a $32\times 4$ cylinder. To transform Eq.~\ref{csf} into momentum space, we introduce
\begin{equation}
c_{\mathbf{k}\sigma}=\frac{1}{\sqrt{N}} \sum_{j} e^{-\,i\,\mathbf{k}\cdot\mathbf{r}_{j}}\,c_{j\sigma}
\qquad(\text{with }N = L W\text{ sites})
\end{equation}
so that
\begin{equation}
\hat{n}_{l} =
\sum_{\sigma} c^{\dagger}_{l\sigma}\,c_{l\sigma}
=
\frac{1}{N}\sum_{\mathbf{k},\,\mathbf{p},\,\sigma}
e^{\,i\,(\mathbf{k}-\mathbf{p})\cdot\mathbf{r}_{l}}
c_{\mathbf{p}\sigma}^{\dagger}\,c_{\mathbf{k}\sigma}.
\end{equation}

Define the Fourier transform of the density fluctuation,
\begin{equation}
\delta\hat n_{\mathbf q} \equiv
\sum_{l} e^{\,i\,\mathbf q\cdot \mathbf r_{l}}\;(\hat n_{l} - n) =
\sum_{\mathbf{k},\,\sigma}
c^{\dagger}_{\mathbf{k}+\mathbf q,\sigma}\,c_{\mathbf{k}\sigma} - n N \delta_{\mathbf q,\mathbf0},
\label{dens_fluct}
\end{equation}
where the geometric sum $\sum_{l}e^{\,i\,(\mathbf q + \mathbf{k} - \mathbf{p})\cdot\mathbf r_{l}}
\;=\; N\,\delta_{\mathbf q + \mathbf{k} - \mathbf{p},\,\mathbf0}$
forces \(\mathbf{p} = \mathbf{k} + \mathbf q\), and \(\delta_{\mathbf q,\mathbf0}\) is the Kronecker delta in momentum space.  Equation~\eqref{csf} can now be written compactly as
\begin{equation}
S_{c}(\mathbf{q}) = \frac{1}{N}\,\bigl\langle \delta\hat n^{\dagger}_{\mathbf q}\,\delta\hat n_{\mathbf q}\bigr\rangle
\label{csfq}
\end{equation}
For a non‐interacting system, the many‐body state is Gaussian, so Wick’s theorem reduces any four‐fermion average to products of two‐fermion ones. With $f_{\mathbf k}  = [e^{\beta(\varepsilon_{\mathbf k}-\mu)}+1]^{-1}$
and
$\varepsilon_{\mathbf k}=-2t(\cos k_x+\cos k_y)$,
\[
\bigl\langle
  \hat c^{\dagger}_{\mathbf k\sigma}\hat c_{\mathbf k+\mathbf q,\sigma}\,
  \hat c^{\dagger}_{\mathbf p\sigma'}\hat c_{\mathbf p-\mathbf q,\sigma'}
\bigr\rangle_{\!\mathrm{GC}}
=\delta_{\mathbf k,\mathbf p-\mathbf q}\,\delta_{\sigma\sigma'}\;
 f_{\mathbf k}(1-f_{\mathbf k+\mathbf q}).
\]
Inserting this into~\eqref{csfq}, summing over the two spin projections,
and noting that the $-nN\delta_{\mathbf q,\mathbf{0}}$ term in
\eqref{dens_fluct} cancels \emph{only} for $\mathbf q=\mathbf{0}$, one obtains
\begin{equation}
S_{c}^{\mathrm{GC}}(\mathbf q)=\frac{2}{N}\sum_{\mathbf{k}}f_{\mathbf{k}}\bigl[1-f_{\mathbf{k}+\mathbf{q}}\bigr].
\label{csfqq}
\end{equation}
The chemical potential $\mu$ is fixed at each $T$
by the single bisection condition
$\frac{2}{LW}\sum_{\mathbf k}f_{\mathbf k}=n$.


\begin{figure}[h]
  \centering
  \includegraphics[width=0.97\textwidth]{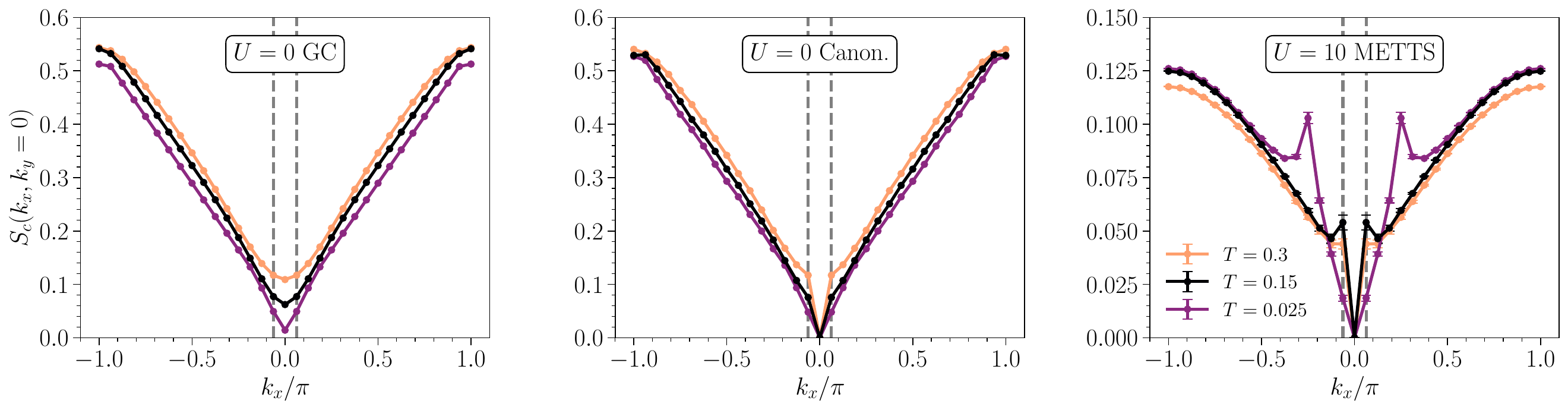}
  \caption{\textbf{Charge Structure Factor --- METTS vs free-fermion comparisons.} Charge–structure factor at $n=0.9375$ for a cylindrical lattice of size $L=32, W=4$.
\textbf{Left:} analytic grand–canonical result  
\textbf{Center:} canonical result from free-fermion snapshots.  
\textbf{Right:} interacting $U=10$ METTS. Only the interacting case exhibits the subtle inner peak at $k_x=2\pi/L$ for intermediate temperatures $T=0.3, 0.15$ and a sharp peak at $k_x=\pi/4$ for the charge-density-wave at $T=0.025$. The presence of the inner peak at intermediate temperatures uniquely in the interacting ($U=10$) canonical METTS data suggests genuine emergent long-wavelength density correlations rather than numerical artifacts or trivial ensemble effects.}
  \label{fig:S6}
\end{figure}


\noindent Next for the charge structure in the fixed-$N$ sector (canonical), we start from the grand-canonical one-body correlator 

\begin{equation}
C_{ij} =
      \bigl\langle \hat c^{\dagger}_j \hat c_i \bigr\rangle_{\mathrm{GC}}
      =
      \sum_{\mathbf k}\varphi_{i\mathbf k}\,
      f_{\mathbf k}\,
      \varphi^{*}_{j\mathbf k},\qquad
      \varphi_{i\mathbf k}=N^{-1/2}e^{i\mathbf k\cdot\mathbf r_i}.
\label{onebodycorr}
\end{equation}

\noindent Because a free system is Gaussian, $C$ alone fixes the density matrix  
$\rho_{\mathrm{GC}}=\mathcal N\exp[-\sum_{ij}(\ln[(1-C)C^{-1}])_{ij}\hat c^{\dagger}_i\hat c_j]$~\cite{peschel2003calculation}. We transform the grand-canonical state into statistically exact $N$-particle Slater determinants by a single measurement sweep:

\begin{enumerate}
\item For site $i$ draw $n_i\in\{0,1\}$ with probability $p_i(1)=C_{ii}$ and $p_i(0)=1-C_{ii}$.
\item Conditionally update the correlator        

\begin{equation}
C\longrightarrow
      \begin{cases}
        C-\dfrac{|v\rangle\langle v|}{p_i(1)}, & n_i=1,\; v_j=C_{ji}\\[6pt]
        C+\dfrac{|u\rangle\langle u|}{p_i(0)}, & n_i=0,\; u_j=\delta_{ji}-C_{ji}
      \end{cases}
\end{equation}

\item Record the occupation $O_i\equiv n_i$ and continue until all sites are visited.
\item Accept the snapshot $\{O_i\}$ only if $\sum_i O_i=N=nLW$; otherwise reject. 
\end{enumerate}

\begin{figure}[t]
    \centering
    \includegraphics[width=0.5\textwidth]{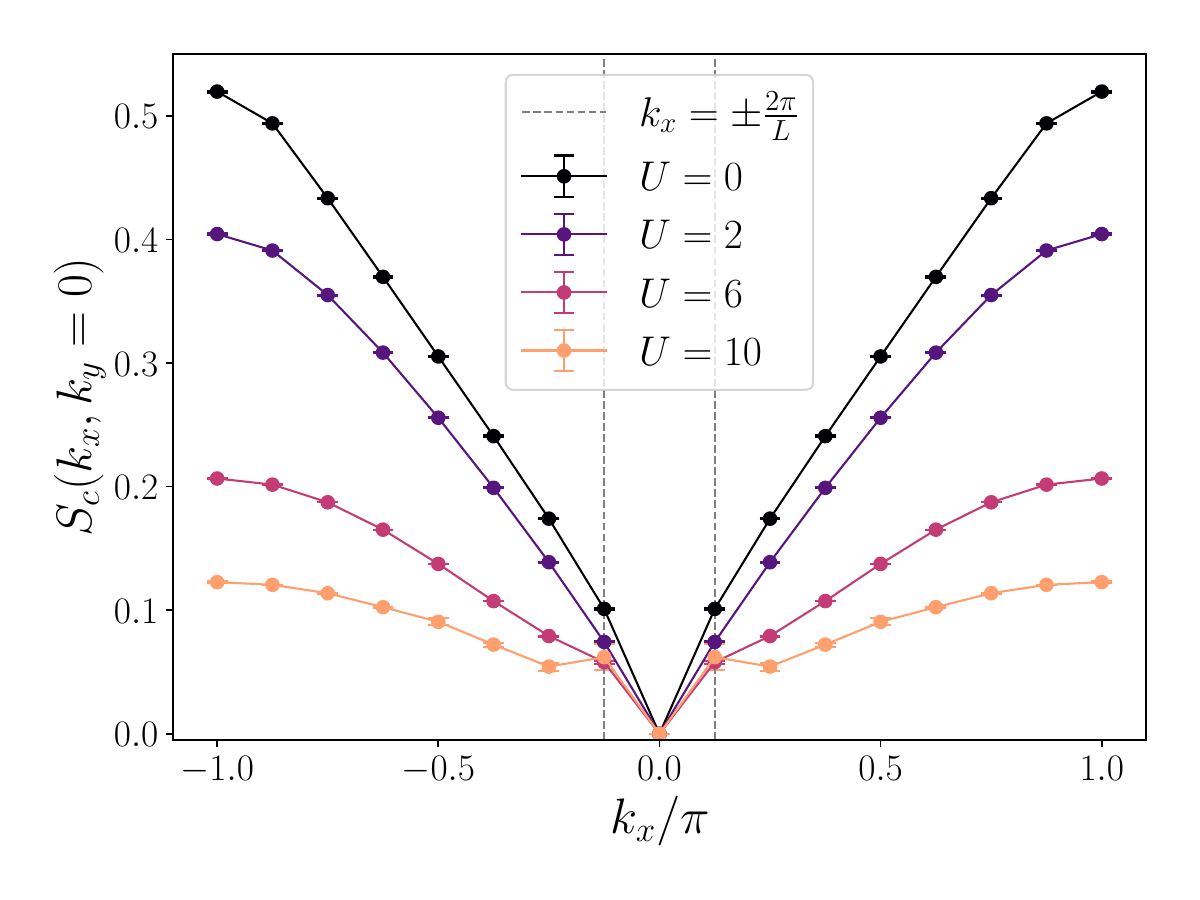}
    \caption{\textbf{Charge Structure Factor for different Interaction Strengths.} Charge structure factor $S_{c}(k_x,k_y=0)$ for a cylindrical lattice of size $L=16, W= 4$ cylinder at $T=0.15$ obtained with METTS in the canonical ensemble. The non-interacting system ($U=0$) shows no inner peak; only at strong coupling ($U=10$) does a clear, though still modest, tiny peak emerge at the smallest non-zero momentum $k_x=\pm2\pi/L$. The emergence of a small peak exclusively at strong coupling ($U=10$) unambiguously connects observed density correlations to interaction-driven physics, rather than trivial finite-size or ensemble effects.}
    \label{fig:S7}
\end{figure}


For every \emph{accepted} snapshot $\alpha$ define the Fourier transform of the density deviation  
\begin{equation}
\widetilde n^{(\alpha)}(\mathbf q)=
    \sum_{j} e^{i\mathbf q\cdot\mathbf r_j}\,\bigl(O^{(\alpha)}_j-n\bigr),
\label{canoncomp}
\end{equation}
which satisfies $\widetilde n^{(\alpha)}(\mathbf0)=0$ exactly because the particle number is fixed.  Since the $O^{(\alpha)}_j$ are ordinary numbers, operator ordering issues are absent and
\begin{equation}
|\widetilde n^{(\alpha)}(\mathbf q)|^{2}
 =\sum_{l,m}e^{i\mathbf q\cdot(\mathbf r_l-\mathbf r_m)}
           (O^{(\alpha)}_l-n)(O^{(\alpha)}_m-n)
\end{equation}
is the desired integrand of the two-point correlator.  Averaging over all accepted snapshots yields the canonical structure factor  
\begin{equation}
S_c^{\mathrm C}(\mathbf q)=
\frac{1}{N}\Bigl\langle
   \bigl|\widetilde n^{(\alpha)}(\mathbf q)\bigr|^{2}
\Bigr\rangle_{\!\alpha}.
\label{}
\end{equation}

Fig.~\ref{fig:S6} gathers the three data sets side by side.  
The grand-canonical curve (left panel, analytic result~\eqref{csfqq}) is
featureless once the trivial $q=0$ weight is excluded, while the
canonical snapshot average (center panel, estimator above) differs from
it only by the expected $\mathcal O(1/N)$ finite-size corrections.  
Indeed, at $T=0.15$ the two free-fermion spectra are already
indistinguishable beyond the first momentum point, and at the lowest temperature $T=0.025$ become almost indistinguishable within plotting resolution.  

In stark contrast, the interacting $U=10$ METTS data (right panel)
display a narrow “inner” peak at $k_x=2\pi/L$ for $T=0.30$ and
$0.15$, a feature that \textit{cannot} be reproduced by either free-fermion
ensemble.  Upon further cooling to $T=0.025$ this weight is transferred
almost entirely to a dominant peak at $k_x=\pi/4$, signaling the
formation of a period-$8$ charge–density wave. Their absence in the free-fermion curves confirms that both peaks are interaction driven.

Finally, we perform a METTS simulations on an $L\times W=16 \times 4$ cylinder at $T=0.15$ for $U=0,2,6,$ and $10$. 
Convergence was reached with bond dimension $D_{\text{max}}=3000$; for $U=0$ the 
resulting $S_{c}$ almost coincides with the direct free-fermion 
snapshots value discussed below. As shown in 
Fig.~\ref{fig:S7}, the $U\le 4$ curves remain essentially undisturbed at $k_x=2\pi/L$, whereas only at $U=10$ does a small peak develop and a barely 
discernible shoulder is present at the intermediate $U=6$ as well.  This demonstrates that canonical‐ensemble weight redistribution alone cannot generate the observed structure–factor peak; its emergence at 
strong coupling reflects genuine long–wavelength density correlations and a tendency towards phase separation.  

\section{Effect of Next–Nearest–Neighbor Hopping ($t'$) on Hole Clustering}
\label{VI}

The square-lattice Hubbard model studied here includes diagonal hopping $t'$ in addition to the nearest-neighbor amplitude $t$:
\begin{equation}
\hat H=
 -t \sum_{\langle i,j\rangle,\sigma}
     \bigl(c^{\dagger}_{i\sigma}c_{j\sigma}+\text{H.c.}\bigr)
 -t' \sum_{\langle\!\langle i,j\rangle\!\rangle,\sigma}
     \bigl(c^{\dagger}_{i\sigma}c_{j\sigma}+\text{H.c.}\bigr)
 +U\sum_i n_{i\uparrow}n_{i\downarrow}.
\label{eq:Hubbard_t_tprime}
\end{equation}

\noindent 

\begin{figure*}[t]
  \centering
  \includegraphics[width=0.9\linewidth]{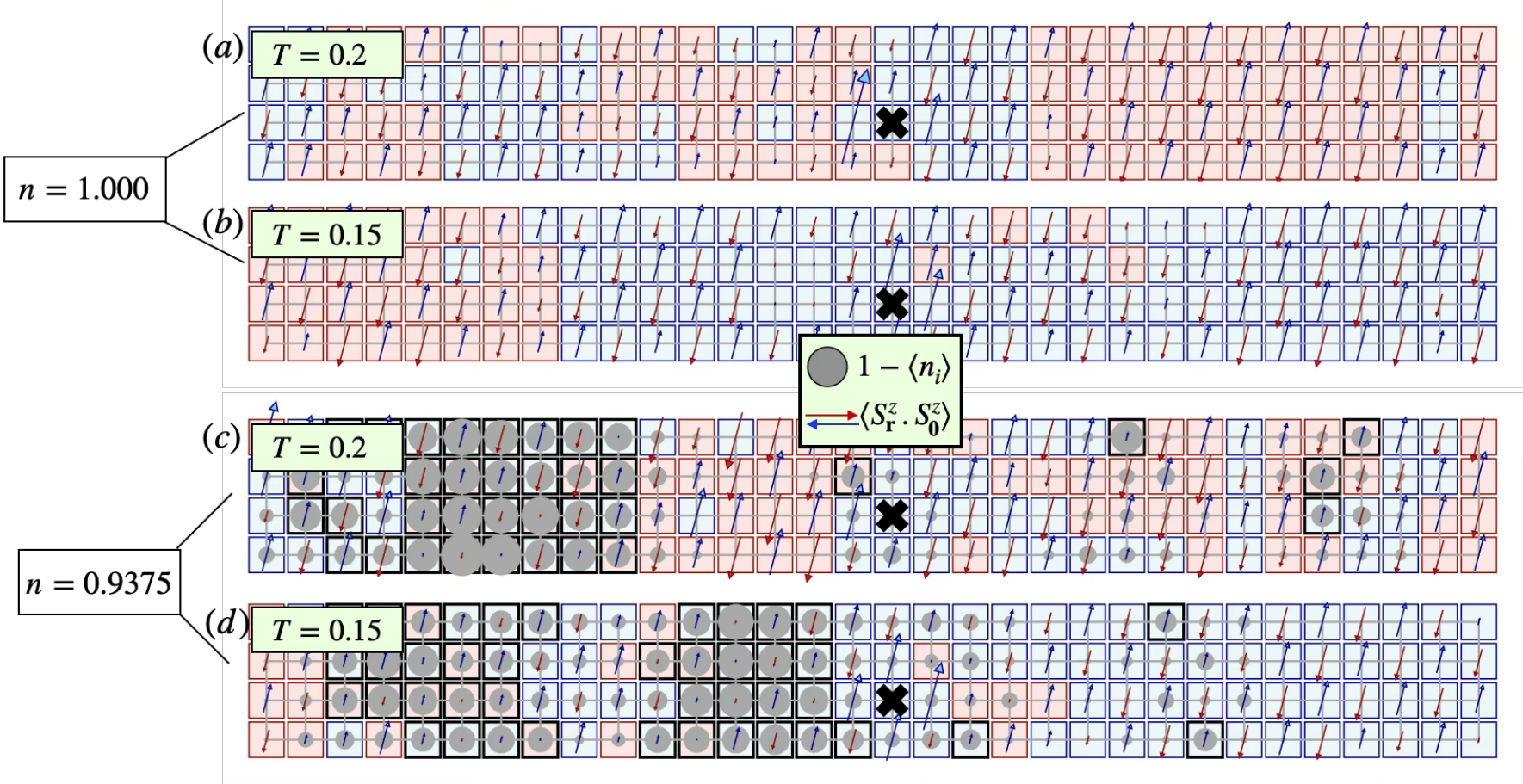}
  \caption{\textbf{METTS snapshots for $t' = 0.3$ Hubbard model.} The typical METTS snapshots are for a cylinder of length $L=32$ and $W=4$, at on-site repulsion $U=10$. The top snapshot at filling $n=1$ for temperatures (a) $T=0.2$ and (b) $T=0.15$ shows no clustering and bottom is at filling $n=0.9375$ for temperatures (c) $T=0.2$ and (d) $T=0.15$ shows hole clustering.  The circle area is proportional to the hole density, arrows show staggered $S^{z}S^{z}$ correlations.  Clustering appears only at finite doping.}
  \label{fig:S8}
\end{figure*}
The results presented in the main text were obtained with zero
next–nearest–neighbor hopping ($t' = 0$) and a lightly doped filling
$n = 0.9375$.  In order to disentangle the roles of kinetic frustration
and mobile holes, we perform METTS simulations at a finite value
$t' = 0.3$ for two electronic densities: the same doped case as before
and, the half-filled Mott insulator $n = 1$.
All calculations were done at interaction strength $U = 10$ on $32 \times 4$ cylinders, a TDVP time
step $\Delta \tau = 0.2$, a maximum bond dimension $D_{\max}=3000$, and
discard the first few METTS iterations to suppress
autocorrelation.  Two temperatures are studied
$T = 0.20$ and $T=0.15$ similar to those in the main text.

Fig.~\ref{fig:S8} contrasts representative METTS snapshots at the
those temperatures.  The lightly doped system at filling $n = 0.9375$ for temperatures (c) $T=0.2$ and (d) $T=0.15$ (bottom row) displays the familiar pattern of hole-rich clusters surrounded by antiferromagnetic domains. The half-filled case
($n = 1$, top row) for temperatures (a) $T=0.2$ and (b) $T=0.15$ remains essentially uniform: staggered spin correlations
alternate regularly, with no sign of segregation. 

\begin{figure}[t]
  \centering
  \includegraphics[width=0.6\linewidth]{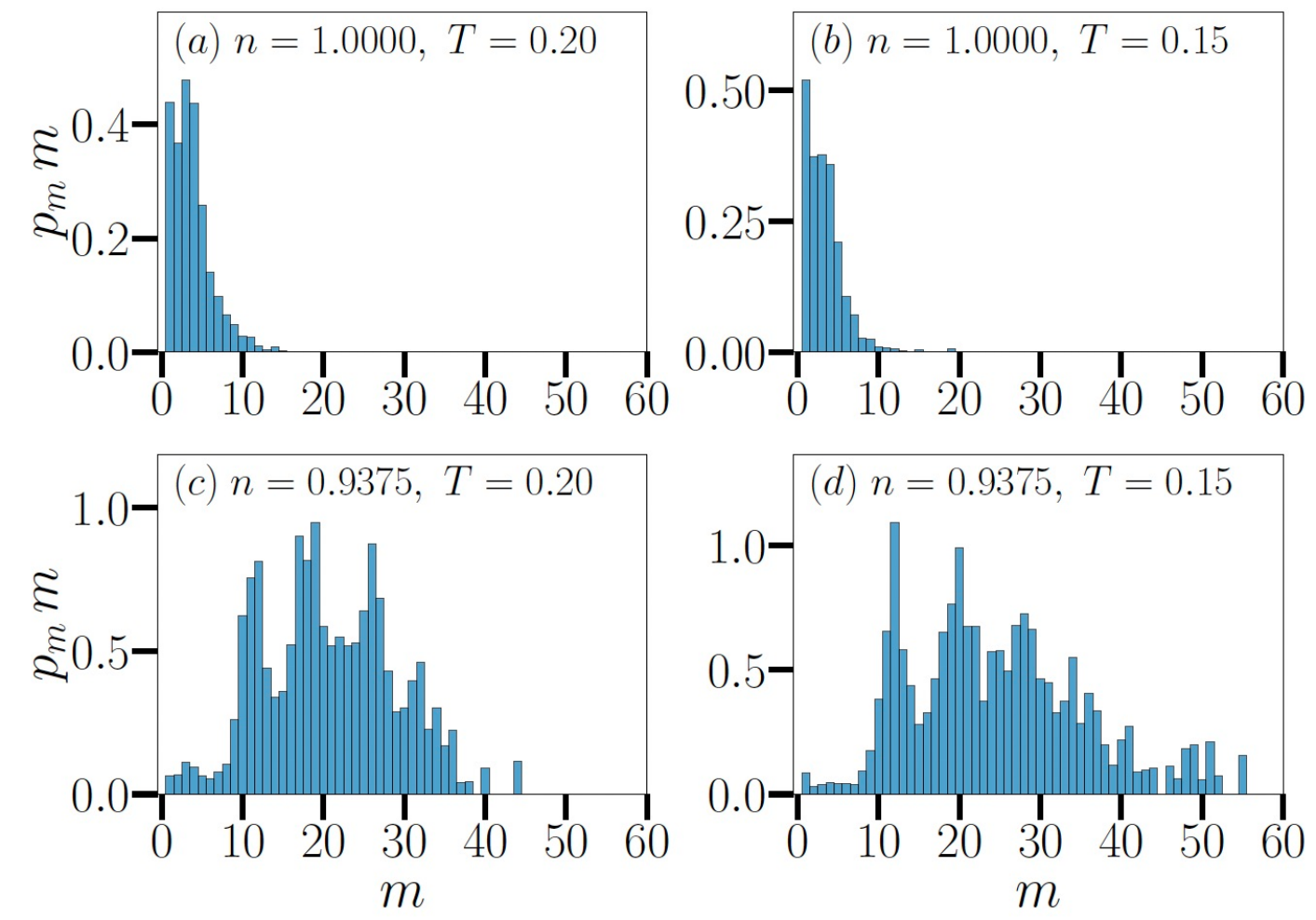}
  \caption{\textbf{Cluster-size statistics for $t'/t = 0.3$ Hubbard model.} Product of  density-weighted probability and cluster size, $p_m m$ vs. cluster size $m$ for filling $n=1$ (temperatures (a) $T=0.2$ and (b) $T=0.15$) and for filling $n=0.9375$ (temperatures (c) $T=0.2$ and (d) $T=0.15$). Broad tails appear only in the doped case. The simulations are conducted on a $L=32$ and $W=4$ cylinder at on-site repulsion $U=10$.}
  \label{fig:S9}
\end{figure}

Cluster–size histograms of $p_{m} m$ vs $m$ for temperatures $T=0.2$ and $=0.15$ in Fig.~\ref{fig:S9} akin to inset of Fig.~\ref{fig:S4} confirm this contrast: at $n=1$ (top row) the distribution is short-ranged, while at $n=0.9375$ (bottom row) it acquires a broad tail. Within the present parameter set, the half-filling remains homogeneous, indicating that kinetic frustration alone is insufficient; mobile holes are required to nucleate charge clusters. A thorough exploration of varying $t'$ and its interplay with doping and temperature will be addressed comprehensively in future work.


\begin{figure}[t!]
  \centering
  \centering
  \includegraphics[width=\textwidth]{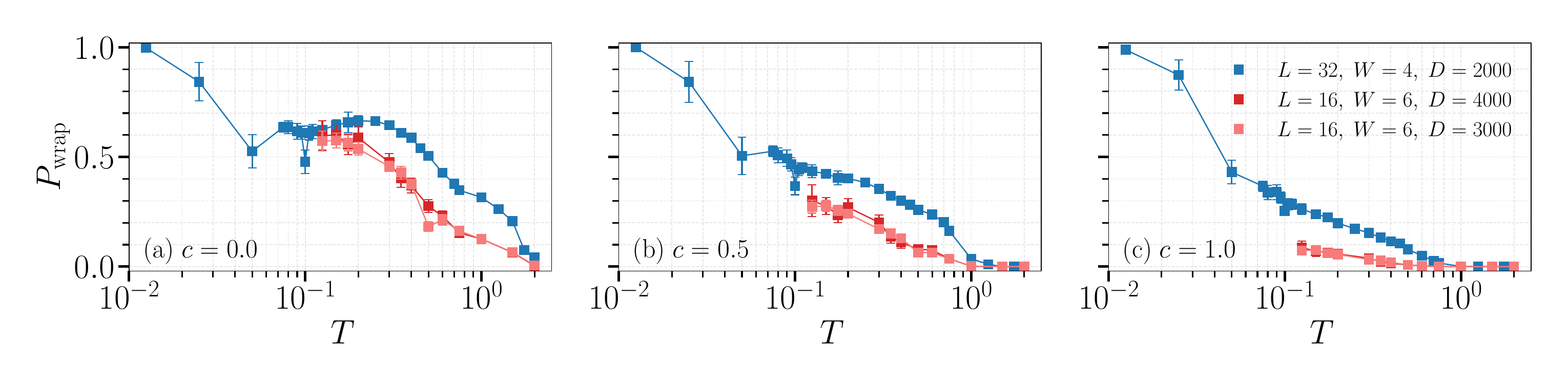}
  \caption{\textbf{Wrap–around probability $P_{\mathrm{wrap}}(T)$ of hole clusters.}  For thresholds $c=0.0,\,0.5,\,1.0$, the fraction of clusters that occupy both transverse edges is plotted versus temperature on a $32\times4$ cylinder at $D=2000$ (blue) and $16\times6$ cylinders at $D=3000$ (orange) and $D=4000$ (green).}
  \label{fig:S10}
\end{figure}
\section{Addressing Finite Size Cylinder Effects}
\label{VII}

\noindent Hole clusters on a finite-width cylinder can, in principle, encircle the periodic transverse boundary and artificially affect the $p_{m}$ distribution. We say that a cluster “wraps” around the cylinder if it contains at least one site on the first transverse edge ($y=0$) and at least one site on the opposite edge ($y=W-1$).  To quantify this, we compute the wrap–around probability
\[
P_{\mathrm{wrap}}(T) = \frac{N_{\mathrm{wrap}}(T)}{N_{\mathrm{tot}}(T)} \,,
\]
where $N_{\mathrm{tot}}(T)$ is the total number of clusters identified at temperature $T$, and $N_{\mathrm{wrap}}(T)$ is the number of those clusters that wrap.

We evaluate $P_{\mathrm{wrap}}(T)$ for threshold coefficients (see Eq.~\ref{th}) $c=0.0,\,0.5,\,1.0$ on $32\times4$ cylinders at maximum bond dimension $D=2000$ and on $16\times6$ cylinders at $D=3000$ and $D=4000$. Fig.~\ref{fig:S10} shows that for all $c$, $P_{\mathrm{wrap}}(T)<0.70$ for $T\gtrsim0.05$ on $W=4$, and $P_{\mathrm{wrap}}(T)\lesssim0.50$ for $T\gtrsim0.10$ on $W=6$. For the $c=0.5$ used in the main text,  $P_{\mathrm{wrap}}(T)<0.50$ for $T\gtrsim0.05$ on $W=4$, and $P_{\mathrm{wrap}}(T)\lesssim0.30$ for $T\gtrsim0.10$ on $W=6$. Thus a non-negligible fraction of clusters do not wrap in the forestalled phase-separation regime. Only below $T\approx0.02$, coinciding with stripe order, does $P_{\mathrm{wrap}}(T)$ rise steeply to unity, consistent with stripe domains spanning the cylinder.

\begin{figure}[t!]
  \centering
  \includegraphics[width=0.95\columnwidth]{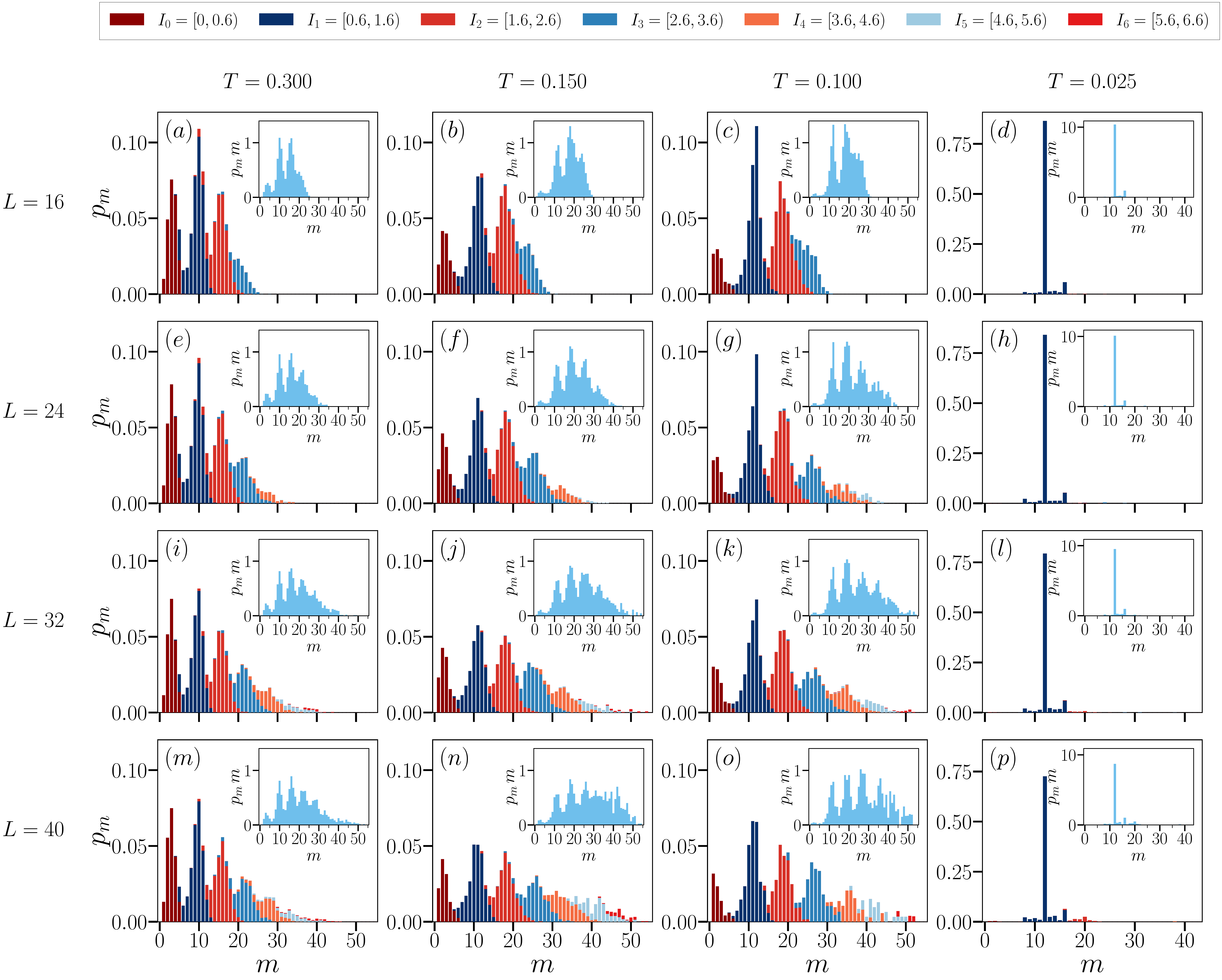}
  \caption{\textbf{Cluster statistics resolved by hole mass.}
  Stacked, hole-mass–resolved $p_m$ vs.\ $m$ for $L=16,24,32,40$ at $n=0.9375$, $U=10$, and $T=0.300,\,0.150,\,0.100,\,0.025$ for fixed width $W=4$. For $T=0.300,0.150,0.100$ the distributions are qualitatively similar (clustering window), with a multi–lobe structure; decreasing $T$ shifts weight slightly to larger $m$.
  Increasing $L$ accommodates larger cluster sizes and reveals additional lobes, while the tail still decays for $m\gtrsim 10$ as expected on finite systems.
  At $T=0.025$ all lengths collapse near $m=12$, consistent with period $\approx 8$ stripe order.}
  \label{fig:S11}
\end{figure}

These findings have two important physical consequences.  First, the smaller wrap–around probability at intermediate temperatures confirms that the observed large clusters reflect genuine, localized hole‐rich domains rather than an artifact of the periodic cylinder. Wrap-around probabilities decrease with width, so the observed hole clustering at intermediate temperatures cannot be attributed to meandering stripes. Such local clustering supports the picture of forestalled phase separation: hole droplets grow with cooling but remain finite until stripe correlations set in.  Second, the sharp crossover to $P_{\mathrm{wrap}}\approx1$ at low $T$ signals the formation of system‐spanning stripes, in agreement with the emergence of long‐range spatial order.

\noindent To disentangle finite length from finite width effects, we scan $L=16,24,32,40$ at fixed $W=4$ for $n=0.9375$ (doping $1-n=1/16$) and $U=10$, using exactly the clustering protocol 
described in the main text. We see in Fig.~\ref{fig:S11}, for $T=0.300,0.150,0.100$ (all within the clustering/forestalled PS window), the distributions are qualitatively similar: $p_m$ has a multi–lobe structure, and lowering $T$ shifts weight slightly toward larger $m$ without changing the overall pattern. Increasing $L$ at fixed $W=4$ increases the total hole number $N_h=(1-n)LW$ 
($N_h=4,6,8,10$ for $L=16,24,32,40$), which 
accommodates larger cluster sizes and reveals 
additional lobes with the same near-integer hole 
mass correspondence. At the same time, the tail of $p_m$ naturally decays for $m\gtrsim 10$ on finite systems, so very large clusters remain rarer than small ones. By contrast, at $T=0.025$ the 
distributions for all $L$ collapse near $m\simeq 12$, consistent with period $\approx 8$ stripe order on $W=4$. Together with the width scan in Fig. $3$ of the main text (where width $W=6$ yields a 
broader $p_m$ than $W=4$ at the same area), these
observations suggest (as a conjecture for the 2D thermodynamic limit) that increasing system extent supports broader distributions and larger cluster sizes while preserving the characteristic lobe pattern associated with near-integer hole aggregation.

%


\bibliographystyle{apsrev4-2}

\end{document}